\documentclass[aps,twocolumn,superscriptaddress,floatfix]{revtex4}

\usepackage{graphicx}

\usepackage{amsmath}
\usepackage{amssymb}
\usepackage{amsthm}
\usepackage{amsfonts}
\usepackage{enumerate}

\usepackage{centernot}
\usepackage{tikz}
\usepackage{ulem}
\usepackage{subfigure}
\usepackage{latexsym}
\usepackage{bm}
\usepackage{color}
\usepackage{txfonts}
\begin{document}

\newcommand{\bq}{{\bf q}}

\title{Intrinsic chiral topological superconductor thin films
}

  \author{Xi Luo}\thanks{These two authors contributed equally.}
  \affiliation{College of Science, University of Shanghai for Science and Technology, Shanghai 200093, People’s Republic of China}

 \author{Yu-Ge Chen}\thanks{These two authors contributed equally.}

  \affiliation{ Institute of Physics, Chinese Academy of Sciences, Beijing 100190, People’s Republic of China}

 \author{Ziqiang Wang}
 \thanks{Correspondence to: wangzi@bc.edu}
 \affiliation{Department of Physics, Boston College, Chestnut Hill, Massachusetts 02467, USA
 }

 \author{ Yue Yu}
 \thanks{Correspondence to: yuyue@fudan.edu.cn}
  \affiliation{State Key Laboratory of Surface Physics, Fudan University, Shanghai 200433,
 	People’s Republic of China}
 \affiliation{Center for Field
 	Theory and Particle Physics, Department of Physics, Fudan University, Shanghai 200433,
 	People’s Republic of China}
 \affiliation{Collaborative Innovation Center of Advanced Microstructures, Nanjing 210093, People’s Republic of China}

\date{\today}

\begin{abstract}
{Superconductors (SCs) with nontrivial topological band structures in the normal state have been discovered recently in bulk materials. When such SCs are made into thin films, quantum tunneling and Cooper pairing take place between the topological surface states (TSSs) on the opposing surfaces. Here, we find that chiral topological superconductivity with spontaneous time-reversal symmetry breaking emerges on the surface of such thin film SCs. There is a  mirror symmetry that protects a novel nonunitary orbital and spin triplet pairing of the TSS.
In the mirror diagonal space, the chiral topological SC manifests as two independent chiral $p$-wave spin-triplet pairing states, in which each is a two-dimensional superconducting analog of the Anderson-Brinkman-Morel state in superfluid $^3$He with in-plane exchange fields. A rich topological phase diagram governed by the nontrivial $\mathbb{Z}\oplus\mathbb{Z}$ topological invariant is obtained with gapless chiral Majorana edge modes and anyonic Majorana vortices.
We further construct a three dimensional  lattice model with a topological band structure and SC pairings,  which is motivated by  Fe-based SCs such as Fe(Te,Se).
We demonstrate the realization of the proposed intrinsic chiral topological superconductor in the quasi-two-dimensional thin film limit.
Our findings enable thin film SCs with nontrivial $Z_2$ band structures as a single-material platform for intrinsic chiral topological superconductivity with both vortex and boundary Majorana excitations for topological quantum device making.
}	
\end{abstract}

\maketitle
\section{Introduction}
Over the past few decades, there has been an intensive search for two-dimensional (2D) chiral triplet superfluids and superconductors (SCs). These are novel quantum states of matter with topological off-diagonal long-range order that support localized non-Abelian Majorana excitations. The latter are the building blocks for topological quantum computation \cite{K1,TQCR}. Candidates for such time reversal symmetry (TRS) breaking topological SCs and superfluids are very {rare} in nature. A widely studied candidate for a chiral $p$-wave superfluid is the Moore-Read state in the $\nu=\frac{5}2$ fractional quantum Hall effect \cite{MR,RG,Stern-usingthisforquantumcomputing}, although experimental evidence is not yet conclusive \cite{will,west}. In bulk crystals with quasi-2D electronic structures, Sr$_2$RuO$_4$ has been proposed to be a chiral $p$-wave SC \cite{rice} and has attracted broad attention to the structure of the spin-triplet pairing order parameter \cite{kallin-review}. However, the experimental evidence for spin-triplet pairing \cite{ishida-nature1998} turned out to be problematic \cite{pustogow-nature2019}. Recently, the proposal \cite{shoucheng1,shoucheng3} for realizing 2D chiral $p$-wave SCs using a quantum anomalous Hall insulator (QAHI) from magnetic topological insulator thin films and SC hybrid structures has been attempted experimentally. The results are highly controversial \cite{science,newscience} and underscore the difficulty in achieving robust chiral topological superconductivity by the proximity effect \cite{JW}.

Electronic structures of crystalline materials
can carry a nontrivial topological ${\mathbb Z}_2$ invariant, such as in 3D strong topological insulators, and support topological surface states (TSSs) \cite{HasanKane,TI-review}. Recently, bulk SCs with ${\mathbb Z}_2$ nontrivial band structures and TSSs in the normal state have been discovered in several Fe-based SCs \cite{arpes-fts,pengzhang-natphys,dinggao2018,hanaguri,donglai-prx,CaKFe4As4,jphu2}.
Candidate Majorana zero modes (MZMs) were observed in magnetic-field-induced vortices \cite{dinggao2018,hanaguri,donglai-prx,CaKFe4As4}
and quantum anomalous vortices \cite{qav} nucleated at the excess and adatom Fe sites in the absence of applied magnetic fields \cite{yin-natphys2015,yin2019,qav-mzm2020} due to the superconducting TSSs \cite{fukane}.
However, these Fe-based SCs are {\it not} ordinary topological superconductors due to the absence of nontrivial topological invariants in the off-diagonal long-range order and boundary Majorana excitations.

\begin{figure}
	\includegraphics[width=0.45\textwidth]{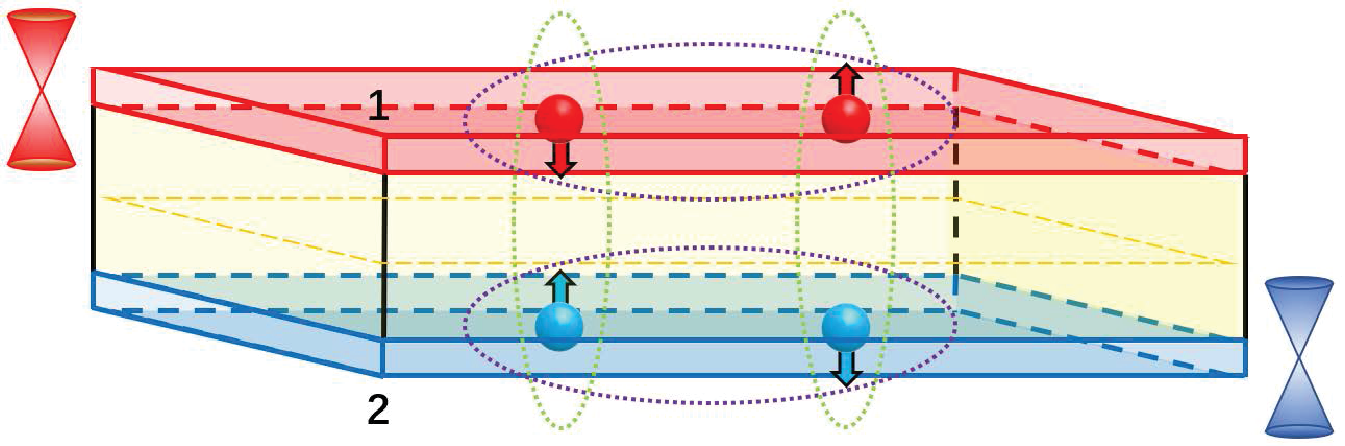}
	\caption{
		Schematics of a $\chi$TSC thin film with coupled superconducting TSS. The  mirror plane is outlined by the brown dashed line with respect to the TSS on the top and bottom surfaces labeled by 1 and 2. The red (blue) balls represent the Dirac fermions localized in the top (bottom) surface, and the arrows represent spin. The pairing terms preserving the mirror symmetry ${\cal M}$ are the spin-singlet pairing on the same surface $\Delta$ (purple dashed line) and the spin-triplet pairing across the surfaces $\Delta_t$  (green dashed line). The quantum tunneling $t$ of the electrons between the surfaces also preserves the mirror symmetry and participates in realizing the $\chi$TSC thin film.
		\label{figX}	}
\end{figure}

Here, we report the discovery of intrinsic chiral topological superconductivity with spontaneous TRS breaking in {\it thin films} of bulk SCs with ${\mathbb Z}_2$ nontrivial band structures where the normal state is time-reversal symmetric. When the top and bottom surfaces of the thin film are brought close and coupled by quantum tunneling and Cooper pairing, the opposite helicities of the Dirac fermion TSS introduce a  mirror symmetry with respect to the $xy$-mirror plane, as illustrated in  Fig.~\ref{figX}. Referring to the two surfaces as two ``orbitals," we find that the pairing interactions of the TSS that preserve the mirror symmetry are the spin-singlet intraorbital triplet and the spin-triplet interorbital singlet described by two independent triplet-pairing {\bf d} vectors. This nonunitary triplet pairing state
spontaneously breaks the TRS and produces a novel chiral topological superconductor ($\chi$TSC) characterized by a nontrivial $\mathbb{Z}\oplus\mathbb{Z}$ topological invariant \cite{sato,mirror-sc}.
The new $\chi$TSC thin films provide topologically protected gapless chiral Majorana edge modes ($\chi$MEM), detectable by the half-quantized conductance at the boundary of an antidot patterned on top of the thin film, in addition to anyonic Majorana vortices.
We demonstrate that thin films of unconventional SCs with nontrivial $Z_2$ invariant band structures and TSSs offer a high-temperature, single-material platform for studying gapless chiral Majorana edge excitations and exploring their applications for topological quantum computing, which may be relevant to Fe-based SCs.

The rest of the paper is organized as follows. We begin with the effective low-energy theory of the superconducting TSS on the thin-film surfaces in Sec.~\ref{secII}. The  mirror-symmetry protected spin-orbital triplet pairings are discussed and shown to produce the nonunitary $\chi$TSC with spontaneous TRS breaking characterized by the topological  $\mathbb{Z}\oplus\mathbb{Z}$ invariant. The model Hamiltonian simultaneously diagonalized in the mirror eigenspace is shown to be equivalent to two independent chiral spin-triplet $p$-wave pairing states,
each of which is a 2D analog of the Anderson-Brinkman-Morel state in superfluid $^3$He under in-plane exchange fields. In Sec.~\ref{secIII}, we study the phase structure of the $\mathbb{Z}\oplus\mathbb{Z}$ $\chi$TSC and present the  topological phase diagrams of the effective Hamiltonian.
The topological phase diagram is extended in Sec.~\ref{secIV} to include the effects of
Zeeman coupling due to an external magnetic field or an incipient magnetic order.
The difference and the robustness of the $\chi$TSC thin films as a platform for $\chi$MEMs are discussed in comparison to the QAHI-SC hybrid structures.
The stability of the thin-film $\chi$-TSC against the substrate potential and pairing interactions that break the mirror symmetry is studied in Sec.~\ref{secV}.
In Sec.~\ref{secVI}, we construct a 3D lattice model for the $Z_2$ nontrivial electronic structure, which is motivated by
Fe-based superconductor Fe(Te,Se), and
{demonstrate the emergence of the intrinsic $\chi$-TSC in the quasi-two-dimensional thin-film limit.}
We also compare and contrast our findings to
{the recent proposal of time-reversal invariant topological superconductivity with helical MEMs in thin films of Fe-based SCs.
\cite{das}.}
In Sec. \ref{secVII}, we study the gapless $\chi$MEMs in the $\chi$TSC phase with nontrivial bulk Chern numbers and propose their experimental detection by half-integer quantized conductance along the boundaries of an antidot patterned on top of the thin film.
Summary and discussions are given in Sec. \ref{secVIII}.


\section{Model and symmetry analysis \label{secII}}


We begin with the low-energy effective theory describing the superconducting TSS on the surface of the thin-film SC illustrated in Fig.~\ref{figX}. The electron annihilation and creation  operators are
$\Psi_{\bf q}=(\psi_{1\uparrow {\bf q}},\psi_{1\downarrow{\bf {q}}},\psi_{2 \uparrow{\bf {q}}},\psi_{2 \downarrow{\bf {q}}})$ and $\Psi_{q}^\dag=(\psi_{1\uparrow {\bf q}}^\dag,\psi_{1\downarrow{\bf {q}}}^\dag,\psi_{2 \uparrow{\bf {q}}}^\dag,\psi_{2 \downarrow{\bf {q}}}^\dag)$, where $1$ and $2$ are the orbital indices labeling the top and bottom surfaces and $\uparrow$ and $\downarrow$ the spin states, and ${\bf q}=(q_x,q_y)$ is the 2D momentum. In the Nambu basis $\Psi_N=(\Psi_{\bf q},\Psi_{\bf -q}^\dag)^T$, we have
$
{\cal H}={1\over2}\sum_{\bf q}\Psi_N^\dagger H({\bf q}) \Psi_N
$ with the $8\times8$ Bogoliubov--de Gennes (BdG) Hamiltonian,
\begin{eqnarray}
H(\bq)&=&
\left(
\begin{array}{cc}
{h}({\bf q}) & {\bf \Delta}^\dag({\bf q}) \\
{\bf\Delta}({\bf q})& -{h}^*({\bf -q}) \\
\end{array}
\right).\label{hbdg}
\end{eqnarray}
The diagonal elements are the normal state Hamiltonian of coupled TSS on the top and bottom surfaces,
\begin{equation}
{h}({\bf q})=
v_F{\bf q}\cdot {\boldsymbol\sigma}\chi_z-\mu\sigma_0\chi_0 +t\sigma_0\chi_x,
\label{h0}
\end{equation}
where $v_F\equiv1$ is the Fermi velocity, $\mu$ the chemical potential, and $t$ the inter-surface tunneling (Fig.~\ref{figX}).
$(\boldsymbol{\sigma},\sigma_0)$ and $({\boldsymbol\chi},\chi_0)$ denote the Pauli and unit matrices acting in the spin and orbital/surface sectors, respectively. The off-diagonal elements ${\bf\Delta}^\dag$ and ${\bf \Delta}$ in Eq.~(\ref{hbdg}) are $4\times4$ pairing matrices to be specified below.

It is crucial to note that due to the opposite helicities of the Dirac fermion TSSs, the normal state Hamiltonian $h({\bf q})$ in Eq.~(\ref{h0}) has a  mirror symmetry $\mathcal{M}$ with respect to the $xy$ plane in the middle of the thin film (Fig.~\ref{figX}), namely, $\mathcal{M} {h}({\bf q}) \mathcal{M}^{-1}={h}({\bf q})$ with
	\begin{equation}
	\mathcal{M}=-i\sigma_z\chi_x, \quad \mathcal{M}^2=-1.
\label{mirror}
	\end{equation}
This mirror symmetry can be used to study the pairing interactions. For isotropic pairing in momentum space, the Cooper pairs must be either a spin singlet and orbital triplet, or a spin triplet and orbital singlet. The superconducting pairing order parameter ${\bf \Delta}$ for the TSS  therefore has the general form
\begin{equation}
	{\bf \Delta}=\Delta (i\sigma_y)({\bf d^\prime}\cdot{\boldsymbol\chi}) (i\chi_y)+\Delta_t ({\bf d}\cdot{\boldsymbol\sigma})(i\sigma_y) (i\chi_y),
\label{delta}
\end{equation}
where ${\bf d}$ and ${\bf d^\prime}$ are two unit triplet $d$ vectors in the spin and orbital sectors. Under the mirror symmetry, ${\cal M}{\bf \Delta}{\cal M}^{-1}={\bf \Delta}$, the only possible $d$ vectors are given by ${\bf d}=\hat z$ and ${\bf d^\prime}=\hat x$, leading to
\begin{equation}
{\bf\Delta}=
i\Delta\sigma_y\chi_z-i\Delta_t\sigma_x\chi_y,
\label{hsc}
\end{equation}
which corresponds to spin-singlet, intraorbital triplet pairing $\Delta$ on the same surface and spin-triplet, interorbital singlet pairing across the two surfaces $\Delta_t$ as illustrated in Fig.~\ref{figX}  \cite{inter1,inter2,inter3}.
This novel isotropic spin-orbital triplet pairing state satisfies
\begin{equation}
{\bf \Delta}{\bf \Delta}^\dag=\vert\Delta\vert^2+\vert\Delta_t\vert^2-i(\Delta_t\Delta^*+h.c.){\cal M},
\label{nonunitary}
\end{equation}
where ${\cal M}$ is precisely the mirror operator in Eq.~(\ref{mirror}). As a result, ${\rm Tr}[{\bf\Delta \Delta}^\dag ({\boldsymbol \sigma}\times {\boldsymbol \chi})]\neq0$, indicating that the pairing state is nonunitary and carries an intrinsic spin-orbit polarization in the mirror eigenspace.

The thin film SC in Eq.~(\ref{hbdg}) is thus described by
\begin{eqnarray}
	H(\bq)&=&  q_x\sigma_x\chi_z\tau_0+q_y\sigma_y\chi_z\tau_z-\mu\sigma_0\chi_0\tau_z
\nonumber \\ &&+t\sigma_0\chi_x\tau_z-\Delta\sigma_y\chi_z\tau_y-\Delta_t\sigma_x\chi_y\tau_y, \label{bdg}
\end{eqnarray}
where $(\boldsymbol{\tau},\tau_0)$ denote the Pauli and unit matrix acting in the particle-hole sector. $H({\bf q})$ has the usual particle-hole symmetry $\mathcal{C}H({\bf q})\mathcal{C}^{-1}=-H(-{\bf q})$, where $\mathcal{C}=\sigma_0\chi_0\tau_x\mathcal{K}$ and $\mathcal{K}$ is the complex conjugation. However, the TRS of the normal state, i.e.,
$\mathcal{T}h({\bf q})\mathcal{T}^{-1}=h({-\bf q})$ with
$\mathcal{T}=i\sigma_y\chi_0\tau_0\mathcal{K}$ and $\mathcal{T}^2=-1$, is broken in the superconducting state in Eq.~(\ref{bdg}) and cannot be restored by crystalline operations {unless one of  $t$, $\Delta$, and $\Delta_t$ is zero. Furthermore, the spin-triplet pairing $\Delta_t$ can possibly be induced from the interlayer electron-electron interactions \cite{nagaosa-prl}.} The thin-film SC is therefore in a nonunitary pairing state with spontaneous TRS breaking, i.e., a chiral superconductor.  More discussions on the TRS breaking are given in Appendix \ref{app1}.


The mirror symmetry implies that the BdG Hamiltonian (\ref{bdg}) has a generalized mirror symmetry
\begin{equation}
\widetilde{\cal M}={\cal M}\tau_z=-i\sigma_z\chi_x\tau_z, \quad [H({\bf q}),\widetilde{\cal M}]=0.
\label{mirror-t}
\end{equation}
Thus, $H({\bf q})$ can be block diagonalized in the basis that diagonalizes $\widetilde{\cal M}$ where the topological classification and the nature of the boundary excitations become transparent.
Specifically, the unitary transformation $U\widetilde{\mathcal{M}}U^{-1}={\rm diag}(-i,-i,-i,-i,i,i,i,i)$ leads to
\begin{eqnarray}
UH(\bq)U^{-1}&=&
\left(
\begin{array}{cc}
{H_D^+}({\bf q}) & 0 \\
0& {H_D^-}({\bf q}) \\
\end{array}
\right), \label{blockh}
\end{eqnarray}
where
\begin{eqnarray}
	H_D^\pm({\bf q})&=& -t\sigma_0\tau_z \pm \mu\sigma_z\tau_z +(\Delta_t\pm\Delta)\sigma_x\tau_z \nonumber\\
&&+q_x\sigma_x\tau_x\mp q_y\sigma_x\tau_y,
\label{hpm}
\end{eqnarray}
are the Hamiltonians in the subspace with mirror eigenvalues $\mp i$. Here, $\sigma_\mu$ continues to act on the spin, whereas $\tau_\mu$ acts in the space which is a mixing of particle-hole and orbital sectors under $U$.
Since $\{\widetilde{\mathcal{M}},\mathcal{C}\}=0$, each block describes a SC with particle-hole symmetry \cite{sato,mirror-sc}, whereas the TRS is broken. As a result, each of $H_D^\pm$  belongs to class $D$ and is classified by a topological invariant $\mathbb{Z}$ in 2D, corresponding to the Chern number of the occupied hole bands. The topological classification of the thin film SC is therefore $\mathbb{Z}\oplus\mathbb{Z}$.

A closer examination of the mirror diagonal $H_D^\pm$ in Eq.~(\ref{hpm}) shows that they describe two independent odd-parity, spin-triplet pairing states with the $d$ vectors ${\bf d}_\pm={\hat z}(q_x\mp iq_y)$, corresponding to  the chiral $p\pm ip$-wave pairing states with orbital angular momenta along the $z$ axis ($J_z=\mp1$). They are the 2D superconducting analog of the Anderson-Brinkman-Morel state in the A-phase of superfluid $^3$He \cite{leggett-rmp,vollhard-book}, which was also proposed for the quasi-2D unconventional SC Sr$_2$RuO$_4$ \cite{mackenzie-rmp}.
The important difference is the presence of the spontaneous polarizing exchange fields ($\Delta_t\pm\Delta$) in the $x$ direction in the mirror eigenspace in Eq.~(\ref{hpm}).
The property of such a topological mirror SC is rich and diverse since it is characterized by both a total Chern number $N=N_+ +N_-$ and a mirror Chern number $C={1\over2}(N_+-N_-)$ \cite{mirrorc}, where $N_\pm$ are the Chern numbers associated with
$H_D^\pm$, respectively.
The $\chi$TSC thin film with spontaneous TRS breaking is achieved when $N\neq0$, whereas the case where $N=0$ but $C\neq0$ describes a class of topological mirror SCs with or without TRS as discussed below.

\section{Topological phase diagrams \label{secIII}}

\begin{figure}
\begin{minipage}{0.23\textwidth}
	\centerline{\includegraphics[width=1\textwidth]{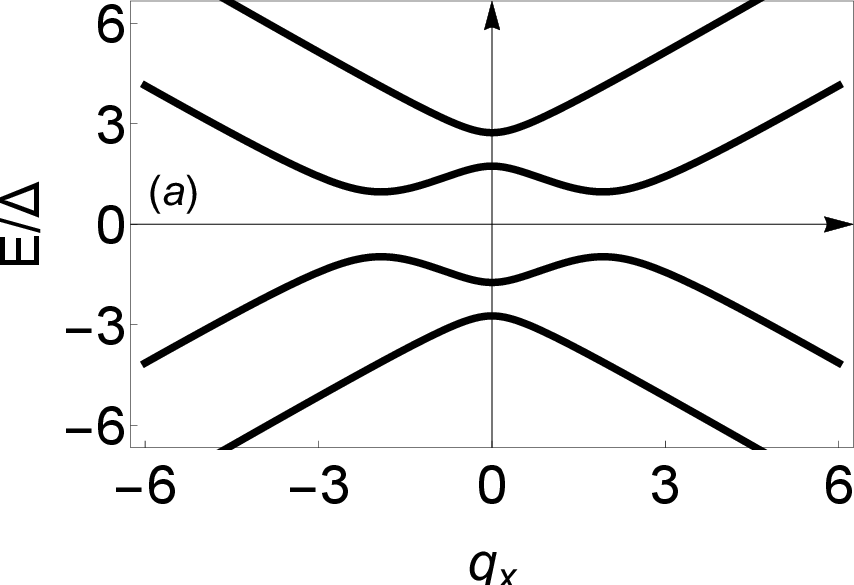}}
\end{minipage}
\begin{minipage}{0.23\textwidth}
	\centerline{\includegraphics[width=1\textwidth]{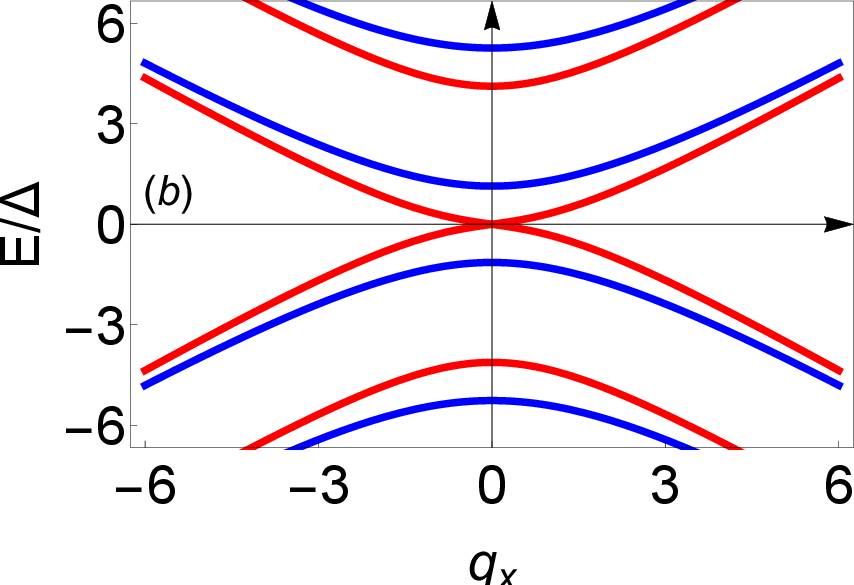}}
\end{minipage}
	 	\begin{minipage}{0.23\textwidth}
	 		\centerline{\includegraphics[width=1\textwidth]{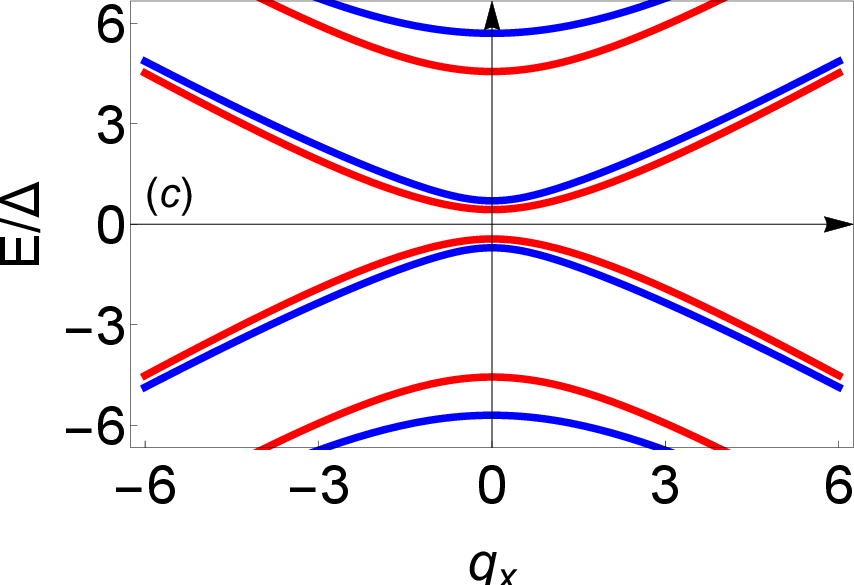}}
	 	\end{minipage}
	 	\begin{minipage}{0.23\textwidth}
	 		\centerline{\includegraphics[width=1\textwidth]{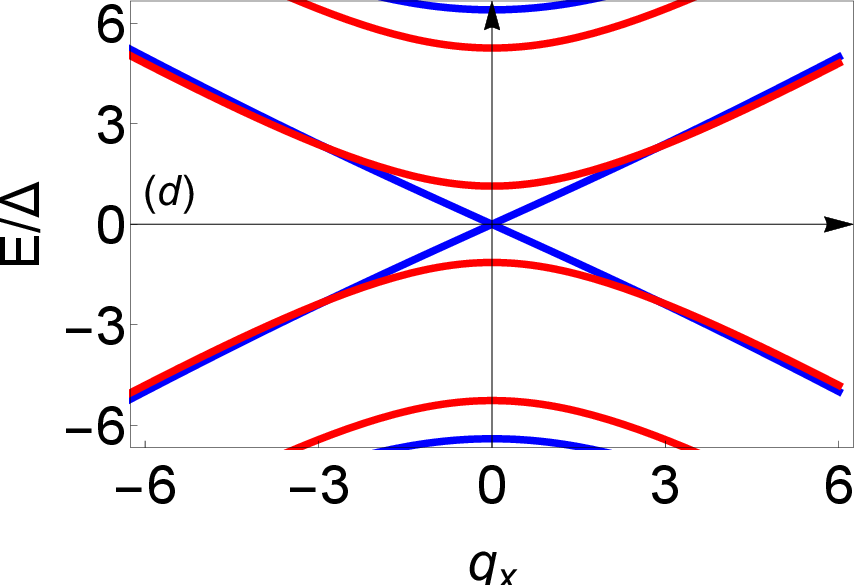}}
	 	\end{minipage}
 	\begin{minipage}{0.23\textwidth}
 		\centerline{\includegraphics[width=1\textwidth]{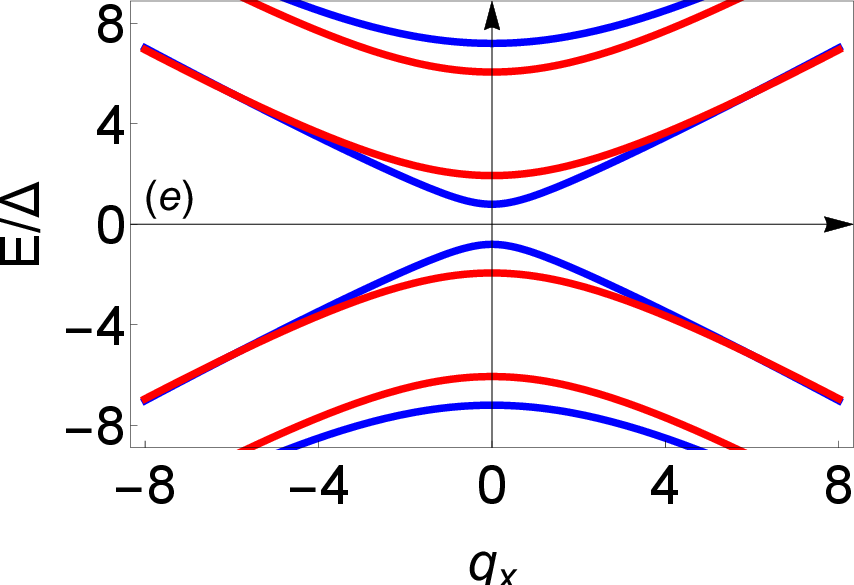}}
 	\end{minipage}
 	\begin{minipage}{0.23\textwidth}
 		\centerline{\includegraphics[width=1\textwidth]{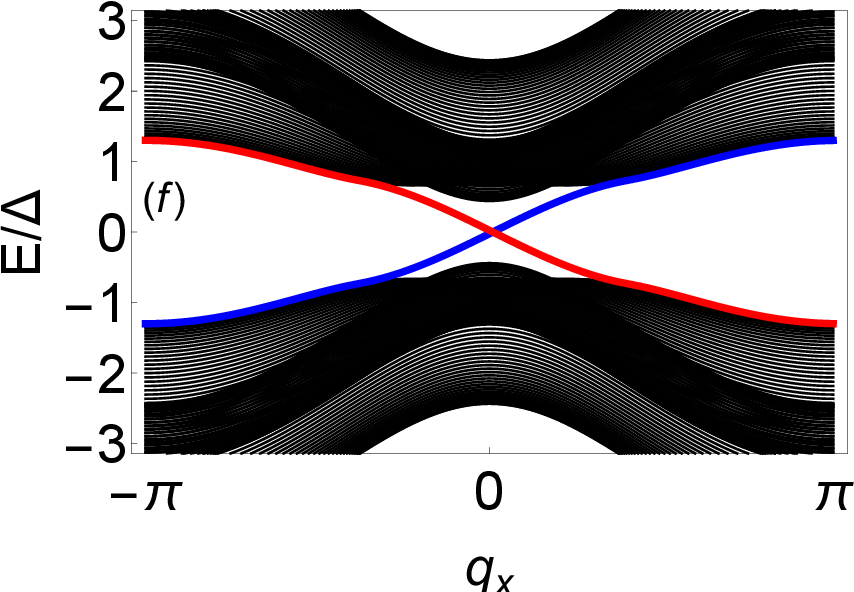}}
 	\end{minipage}
	\caption{
		Energy spectrum of the thin film SC described by the BdG Hamiltonian given by Eq.~(\ref{bdg}), plotted in the $q_y=0$ direction for $\Delta=1$ and $\mu=2$. The spin-triplet, orbital-singlet pairing (a) $\Delta_t=0$ and (b)-(e) $\Delta_t=1.5$. The interorbital tunneling evolves as (a) $t=0.5$, (b) $t=2.06$, (c) $t=2.5$, (d) $t=3.2$, and (e) $t=4$. In (b)-(e), the blue (red) curves correspond to the spectra
of $H_D^+$ ($H_D^-$). (f) Energy spectrum of the lattice model with the same parameters as in (c) under the open boundary condition in the $y$ direction. The red and blue curves are the gapless states of a single $\chi$MEM localized at the two edges.
		\label{fig1}	}
\end{figure}

 The quasiparticle spectrum of the BdG Hamiltonian in Eq.~(\ref{bdg}) can be obtained analytically from those of the diagonal blocks $H_D^{\pm}$. Denoting $\varepsilon_{\pm}^+=[q^2+(\Delta_t+\Delta\pm t)^2]^{1/2}$ and $\varepsilon_{\pm}^-=[q^2+(\Delta_t-\Delta\pm t)^2]^{1/2}$ with $q^2=q_x^2+q_y^2$, they are given by $E_\pm^+=\pm\varepsilon_\pm^+$ and $E_\pm^-=\pm\varepsilon_\pm^-$ at $\mu=0$.
We plot in Fig.~\ref{fig1} the evolution of the eight eigenbands. The values of $\mu$ and $\Delta$ are chosen to be in the range of the experimental values for Fe$_{1+y}$Se$_x$Te$_{1-x}$ with $\Delta/\mu=0.5$ and {$\Delta=2$meV} \cite{dinggao2018,d-mu}. Hereafter, $\Delta$ is set to unity as the energy unit. We begin with the case $\Delta_t=0$ shown in Fig.~\ref{fig1}(a). The nonzero chemical potential offsets the two Dirac cones of the top and bottom surfaces. The two Dirac points separated by $2\mu$ are gapped by the interorbital tunneling $t$. The intraorbital spin-singlet pairing $\Delta$ opens an energy gap at the Fermi level.  Each band in Fig.~\ref{fig1}(a) is doubly degenerate and the SC is topologically trivial.

Turning on ${\Delta_t}\neq0$, the band degeneracy is lifted due to TRS breaking. Increasing $t$ or {$\Delta_t$} reduces the superconducting gap and leads to a gap closing at the $\Gamma$ point in {Fig.~\ref{fig1}(b)}. The gap reopens with a band inversion in {Fig.~\ref{fig1}(c)}, following a topological phase transition to an emergent $\chi$TSC with a total Chern number $N=-1$ [see Fig. \ref{fig3}(b)]. This is  directly confirmed by a lattice model calculation (see {Appendix \ref{app2}}) showing a single Majorana edge mode localized at the open boundaries in Fig.~\ref{fig1}(f).
Further increasing the tunneling $t$ can cause the gap to close again as in Fig.~\ref{fig1}(d) and the system to return to a topologically trivial SC in Fig. \ref{fig1}(e).


To determine the topological phase structure, we numerically calculate the Berry phase of each band.
The total Chern number is given by the sum of the Berry phases \cite{TKNN} of the lower four quasihole bands. In Figs. \ref{fig3}(a)--3(c), the obtained phase diagrams are shown in the $\Delta-\mu$ plane for different values of the interorbital spin-triplet pairing $\Delta_t$. Since the system is a topological crystalline SC protected by the mirror symmetry $\mathcal{\widetilde M}$, the topological region with nontrivial Chern number $N$ is a superposition of that of the diagonal blocks $H_D^\pm$ in each mirror eigenspace. The phase boundaries in Figs. \ref{fig3}(a)--3(c) are thus determined by the condition
\begin{equation}
\mu^2+(\Delta\pm\Delta_t)^2=t^2,
\label{phaseequation}
\end{equation}	
for the band inversion at the $\Gamma$ point in $H_D^\pm$ in Eq.~(\ref{hpm}), respectively. They correspond to the two circles centered at $(\mp\Delta_t,0)$ in the $\Delta-\mu$ plane with the same radius $t$ in Figs. \ref{fig3}(a)--3(c).
As a result, in the absence of tunnel coupling between the top and bottom surfaces, i.e. when $t=0$, the superconducting state is topologically trivial. For a small $t\ne0$, each circle encloses a $\chi$TSC as shown in Fig. \ref{fig3}(a), with the total Chern number $N=N_++N_-=\pm1$, arising from $(N_+,N_-)=(1,0)$ and $(0,-1)$ respectively. There is a single $\chi$MEM [see Fig.~\ref{fig1} (f)] at the boundary of the thin-film SC.

Increasing $t$ or reducing the magnitude of the spin-triplet pairing $\Delta_t$ between the top and bottom surfaces causes the two circles to move toward each other and overlap, as shown in Fig.~\ref{fig3}(b). The phase in the overlapping (red) regime has a total Chern number $N=0$, but a mirror Chern number $C={1\over2}(N_+-N_-)=1$, and is therefore a topological mirror SC.
There are two counter propagating edge modes along a single edge, but backscattering between them is prohibited by the mirror symmetry ${\cal\widetilde M}$.
This topological mirror SC with an odd mirror Chern number $C$ remains stable in the limit $\Delta_t=0$,
where the two $\chi$TSCs in the mirror eigenspace with opposite chirality become degenerate as shown in  Fig. \ref{fig3}(c).  The absence of pairing between the two surfaces restores the TRS (see {Appendix \ref{app1}}) and enables a {$\mathbb{Z}_2\oplus\mathbb{Z}$  topological classification, where $\mathbb{Z}$ labels the mirror Chern number,} producing the helical Majorana fermion edge states at the open boundaries shown in  Fig.~\ref{fig3}(d) in lattice model calculations.
%

{Before ending this section, we would like to comment on the coexistence of $\Delta$ and $\Delta_t$. In general, for a 3D superconductor with inversion symmetry and periodic boundary condition, the bulk superconductivity is either purely spin-singlet pairing or purely triplet pairing, distinguished by their parities. The existence of surface breaks inversion symmetry, and the mixture of these two kinds of pairings is allowed \cite{RPP2017,JPCM2021}. A similar situation appears for the 2D effective theory, where the inversion symmetry is partially broken with a remaining mirror-symmetry operation interchanging the top and bottom surfaces. The square lattice model that we study has a $D_{4h}$ symmetry and, in the thin-film limit, $\Delta$ and  $\Delta_t$ are all parity odd; therefore, it is possible for them to mix and coexist.

{In Appendix C, we present a Ginzburg-Landau free-energy analysis.
Minimizing the Ginzburg-Landau free energy, we find that there are two distinct stable states where $\Delta$ and  $\Delta_t$ coexist with different relative phases between the two odd-parity pairing order parameters. The state with the phase structure $(\Delta,\Delta_t)\propto(\pm1 ,\pm i)$ has a $\pi/2$ relative phase between
$\Delta$ and  $\Delta_t$, leading to a time-reversal invariant unitary state by Eq.~(\ref{nonunitary}). In contrast, the other stable state has the phase structure {$(\Delta,\Delta_t)\propto(\pm1,\pm1)$}, which precisely describes the nonunitary pairing state by Eq.~(\ref{nonunitary}) that we proposed and studied here. It breaks the time-reversal symmetry and is an intrinsic chiral topological superconductor \cite{JPCM2021,AP1990,RMP1991}. Which of these states has lower energy in the Ginzburg-Landau theory depends on the values of the parameters determined  by the microscopic details of the model for specific materials, including both the band structure and electron-electron interactions. Our purpose here is to propose a different mechanism and a proof of principle that it is possible to realize intrinsic chiral topological superconductivity in thin films of superconductors with a topological band structure. The microscopic models based on realistic materials band structures and electron-electron interactions are clearly beyond the scope of the current paper and will be pursued in the future. In the following sections, we continue to study the properties of the intrinsic chiral topological superconducting phase.}
%

\begin{figure}
	\begin{minipage}{0.23\textwidth}
		\centerline{\includegraphics[width=1\textwidth]{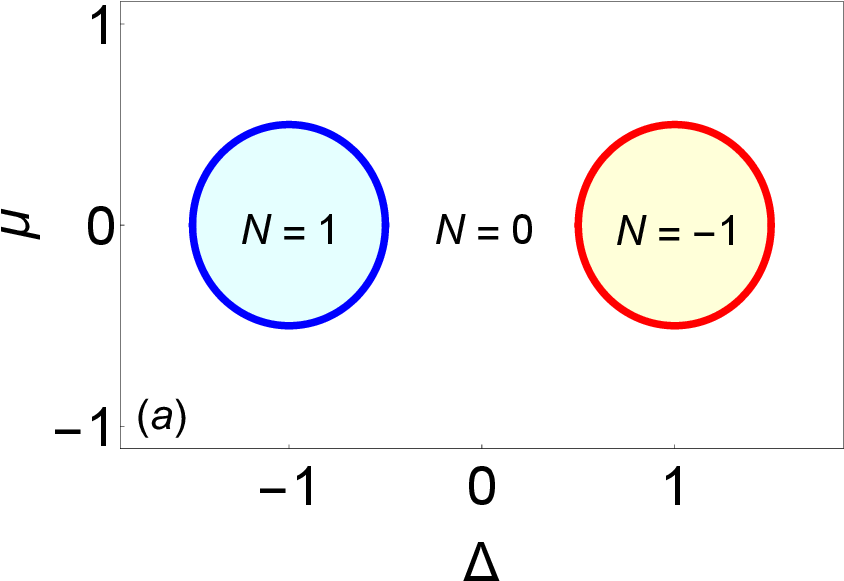}}
	\end{minipage}
	\begin{minipage}{0.23\textwidth}
		\centerline{\includegraphics[width=1\textwidth]{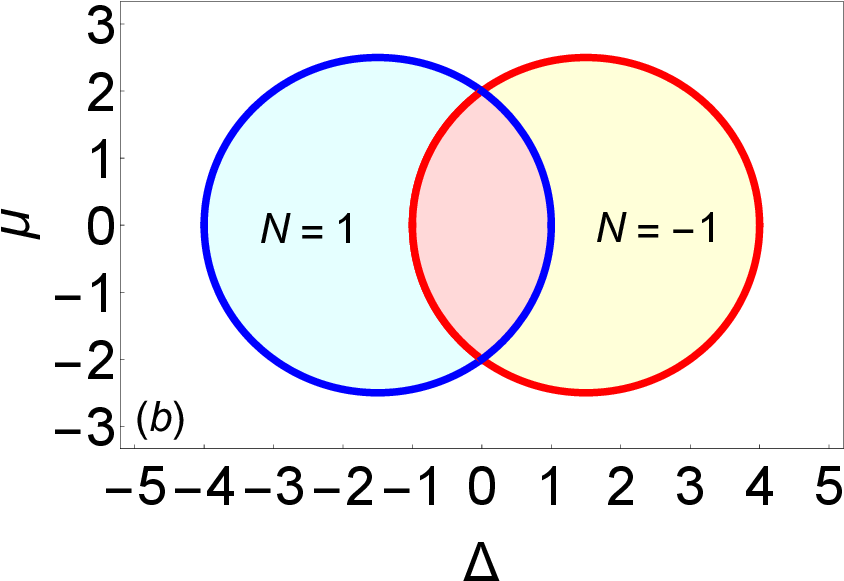}}
	\end{minipage}
	\begin{minipage}{0.23\textwidth}
		\centerline{\includegraphics[width=1\textwidth]{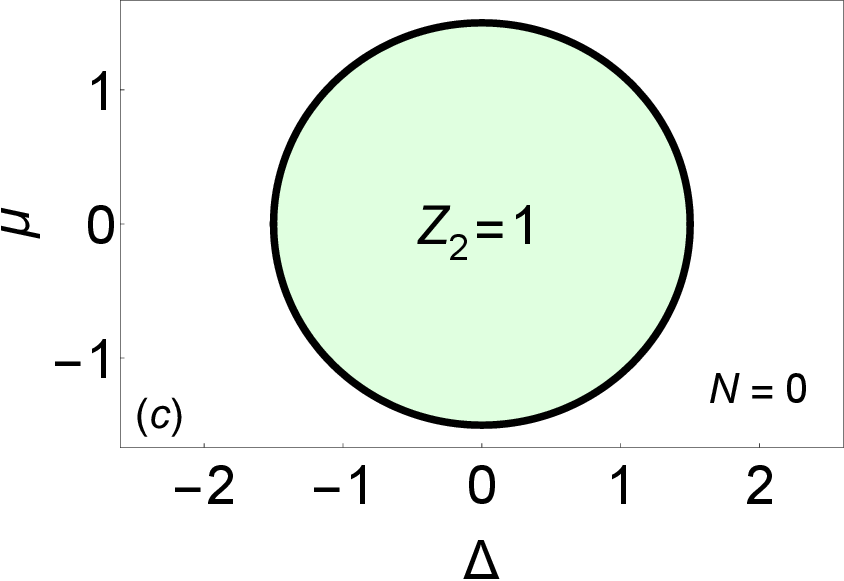}}
	\end{minipage}
	\begin{minipage}{0.23\textwidth}
		\centerline{\includegraphics[width=1\textwidth]{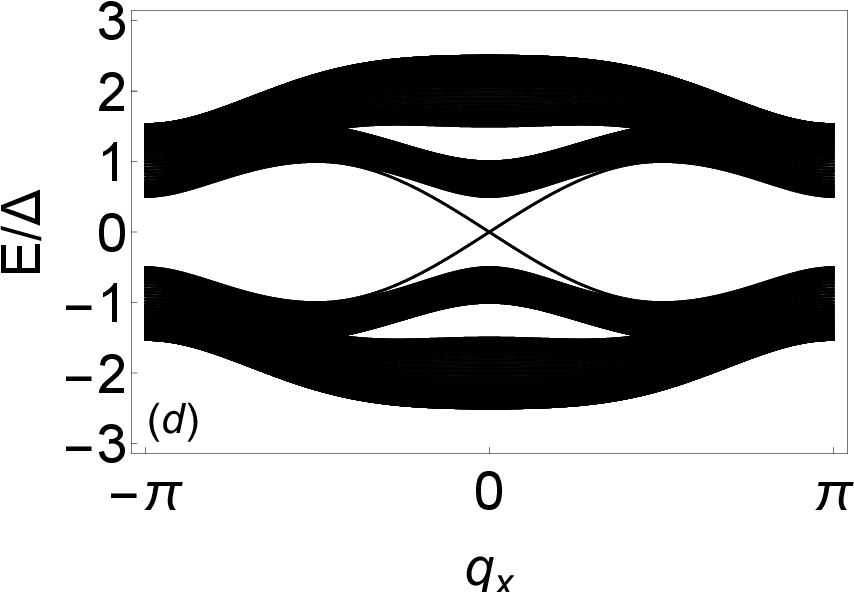}}
	\end{minipage}
	\caption{
		Topological phase diagrams of the thin film SC in the $\Delta$-$\mu$ plane for (a) $t=0.5$, $\Delta_t=1$, (b) $t=2.5$, $\Delta_t=1.5$, and (c) $t=1.5$, $\Delta_t=0$.  The $\chi$TSC phases are labeled by the total Chern number $N\neq0$ originating from {$H_D^+(H_D^-)$} in regions bounded by the blue (red) curves.
		For comparison, the energy spectrum shown in Fig. \ref{fig1}(c) corresponds to the ($\Delta$=1, $\mu=2$) point in (b). The red overlapping region in (b) is a topological nontrivial phase with $N=0$ and mirror Chern number $C=1$, while the green phase region in (c) is characterized by $\mathbb{Z}_2$ due to the restored TRS. (d) The energy spectrum of the lattice model with open boundaries in the $y$ direction, showing the gapless helical edge modes. The results are obtained at the $\Delta$=1, $\mu=0$ point in (c).
		\label{fig3}	}
\end{figure}

\section{Effects of  Zeeman Field\label{secIV}}

 Thus far, our $\chi$TSC thin film spontaneously breaks the TRS by nonunitary pairing. We now generalize the topological phase diagram to include the effects of Zeeman coupling $\lambda$ due to either an applied magnetic field normal to the surfaces or an incipient ferromagnetic order. The Zeeman coupling enters the BdG Hamiltonian in Eq.~(\ref{bdg}) according to $H+\lambda\sigma_z\chi_0\tau_z$, which preserves the mirror symmetry { $\widetilde{\cal M}$}. The block-diagonal Hamiltonian in the mirror eigenspace in Eq.~(\ref{blockh}) becomes $H_D^\pm(t\to t\pm\lambda)$, with corrections to the quantum tunneling $t$ by $\lambda$ in opposite directions. The resulting topological phase diagrams are shown in Fig. \ref{fig-lambda} in the $\Delta_t-\lambda$ plane for two different values of $t$. Along the $\lambda=0$ line, the TRS is spontaneously broken by {$t$ and the nonunitary} pairing, and produces the $N=\pm1$ {\it intrinsic} $\chi$TSC with a single $\chi$MEM. The half-quantized conductance $\frac{e^2}{2h}$ of the $\chi$MEM can be directly detected at the edge of an antidot, which will be discussed below.

\begin{figure}
	\begin{minipage}{0.23\textwidth}
		\centerline{\includegraphics[width=1\textwidth]{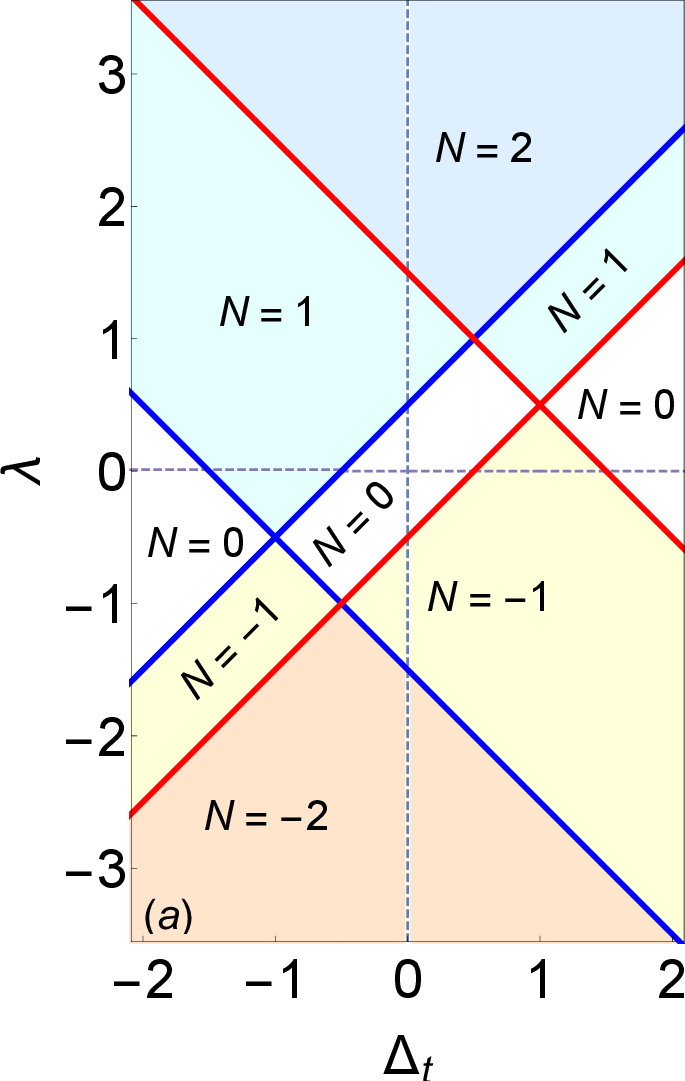}}
	\end{minipage}
	\begin{minipage}{0.23\textwidth}
		\centerline{\includegraphics[width=1\textwidth]{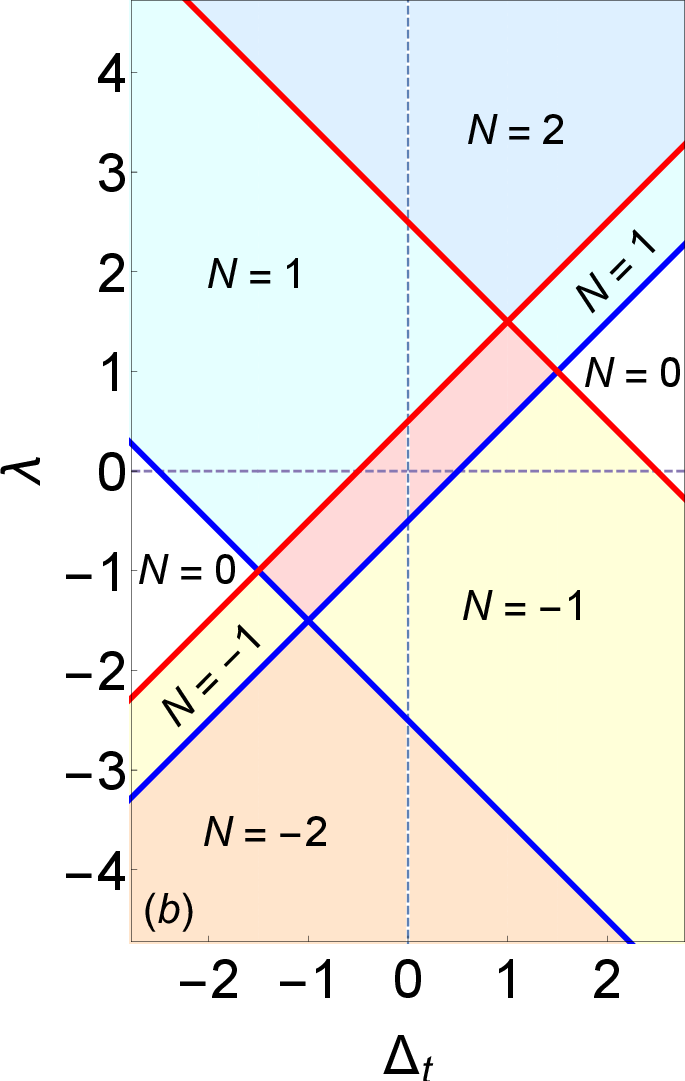}}
	\end{minipage}
	\caption{
		Topological phase diagrams in the $\Delta_t-\lambda$ plane for $\Delta=1$, $\mu=0$, and (a) $t=0.5$ and (b) $t=1.5$. The $\chi$TSC phases are labeled by the total Chern number $N\neq0$ originating from {$H_D^+(H_D^-)$} in the regions bounded by the blue (red) curves. The rectangle enclosing the origin in (b) is a mirror TSC with $N=0$ and $C=1$.
		\label{fig-lambda}}
\end{figure}

Extending the phase diagrams into the regions where the TRS is also explicitly broken by $\lambda\neq0$, we find that most of the phases are $\chi$TSCs, including additional higher Chern number ($N=\pm2$) phases. Note that since $\lambda$ preserves the mirror symmetry, a mirror index $C$ emerges in the rectangular region surrounding the origin in Fig. \ref{fig-lambda}. A nontrivial $C=1$ mirror TSC in Fig. \ref{fig-lambda}(b) arises due to the larger tunnel coupling between the top and bottom surfaces. The $\chi$TSC phases along the $\Delta_t=0$ line in Fig.~\ref{fig-lambda} are formally analogous to the proposed QAHI-SC proximity effect hybrid structures  {\cite{shoucheng1,science,hih}}, where doped magnetic-ion-induced ferromagnetism (thus $\lambda$) in a topological insulator thin film is crucial for breaking the TRS.
However, the hybrid structure may end up in a metallic phase instead of the desired $\chi$TSC \cite{JW}. There may be magnetic-ion-induced vortices trapping MZMs \cite{qav} that dramatically change the results of non-Abelian braiding \cite{nonabelian1,nonabelian2} in the proposed QAHI-(QAHI+SC)-QAHI device \cite{shoucheng3}. These obstacles are absent in the intrinsic $\chi$TSC thin films proposed here. Without the complications of the Zeeman or exchange field and the superconducting proximity effect, the resulting $\chi$MEMs can provide a more robust and advantageous platform for constructing the topological logic gates for quantum computing \cite{Fib-Eps}.



\section{Mirror symmetry breaking effects \label{secV}}

In realistic materials realizations of the thin film, substrates and the bulk superconductivity can induce substrate potentials and pairing interactions between the top and bottom TSSs that break the  mirror symmetry. In this section, we study these effects and demonstrate the stability of the thin-film $\chi$TSC against mirror-symmetry breaking due to the protection of the topological gap. {The mirror-symmetry protection of $\chi$TSC is similar to that of topological insulator (TI), namely, TI is protected by TRS, and is stable against small magnetic perturbations as long as the topological gap remains open. }

\subsection{Mirror symmetry breaking pairing}

To study the effects of mirror-symmetry breaking pairing, we can rotate the spin-singlet, orbital-triplet pairing ${\mathbf d^\prime}$ vector in Eq.~(\ref{delta}) away from the ${\hat x}$ axis into the $xz$ plane. Physically, this corresponds to adding to Eq.~(\ref{delta}) the spin-singlet pairing between the top and bottom surfaces described by $\Delta_s(i\sigma_y) ({\mathbf d^{\prime\prime}}\cdot{\boldsymbol \chi})(i\chi_y)$ with ${\mathbf d^{\prime\prime}}={\hat z}$. The resulting pairing function in Eq.~(\ref{hsc}) becomes
\begin{equation}
{\mathbf \Delta}=
i\Delta\sigma_y\chi_z-i\Delta_t\sigma_x\chi_y-i\Delta_s\sigma_y\chi_x.
\label{hsc2}
\end{equation}
Since the mirror symmetry $\widetilde{\mathcal{M}}$ in Eq.~(\ref{mirror-t}) is broken, the modified BdG Hamiltonian in Eq.~(\ref{bdg}) with $\Delta_s\neq0$ can no longer be block diagonalized. However, it continues to describe a nonunitary pairing state due to the broken TRS. We thus directly obtain the topological phase diagrams in the $\Delta_t-\Delta_s$ plane by numerical diagonalization. Figures~\ref{nu2}(a) and \ref{nu2}(b) show
that the chiral topological superconducting phases with nontrivial Chern numbers $N=1$ and $N=-1$ remain stable. They are protected by the robust topological gap and separated from the topological trivial phases marked by $N=0$ by gap-closing transitions. Due to the broken mirror symmetry, the thin film $\chi$TSC now belongs to class $D$ and is classified by the topological invariant $\mathbb{Z}$.

\begin{figure}
	\begin{minipage}{0.23\textwidth}
		\centerline{\includegraphics[width=1\textwidth]{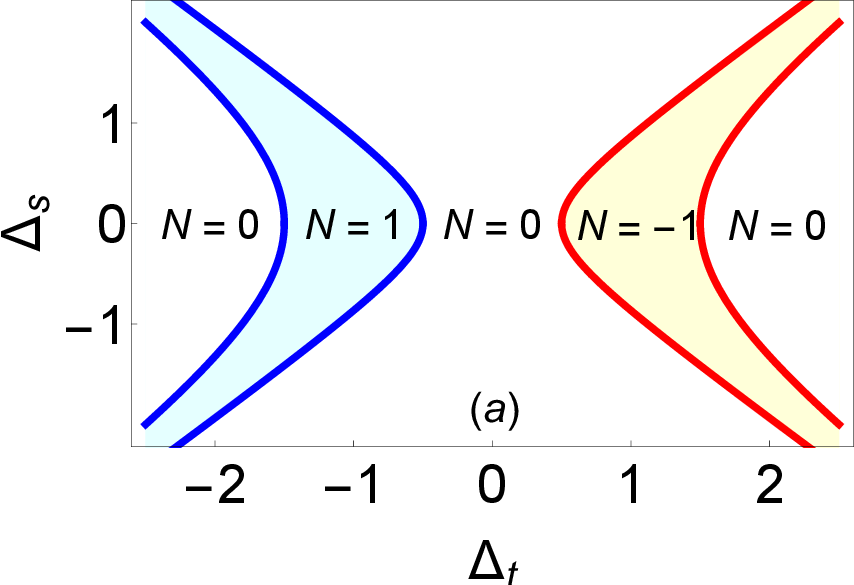}}
	\end{minipage}
	\begin{minipage}{0.23\textwidth}
		\centerline{\includegraphics[width=1\textwidth]{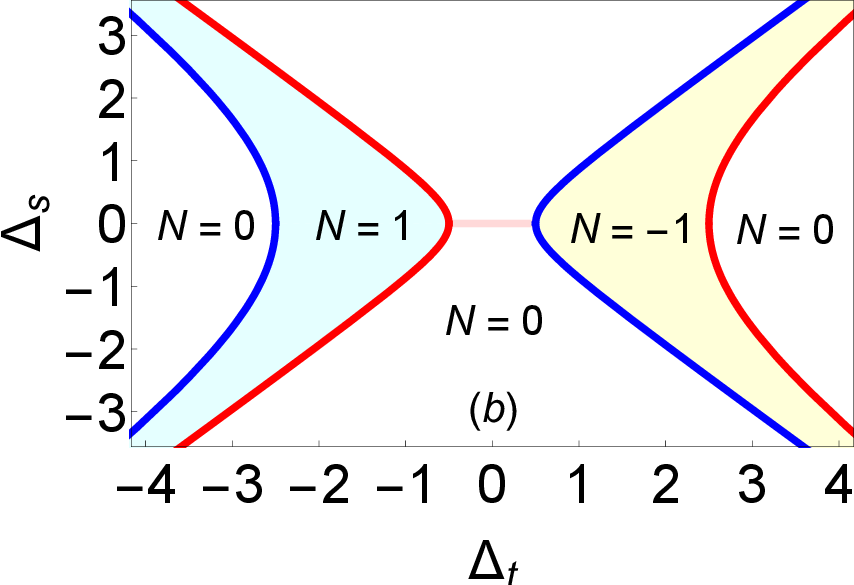}}
	\end{minipage}
	\begin{minipage}{0.23\textwidth}
	\centerline{\includegraphics[width=1\textwidth]{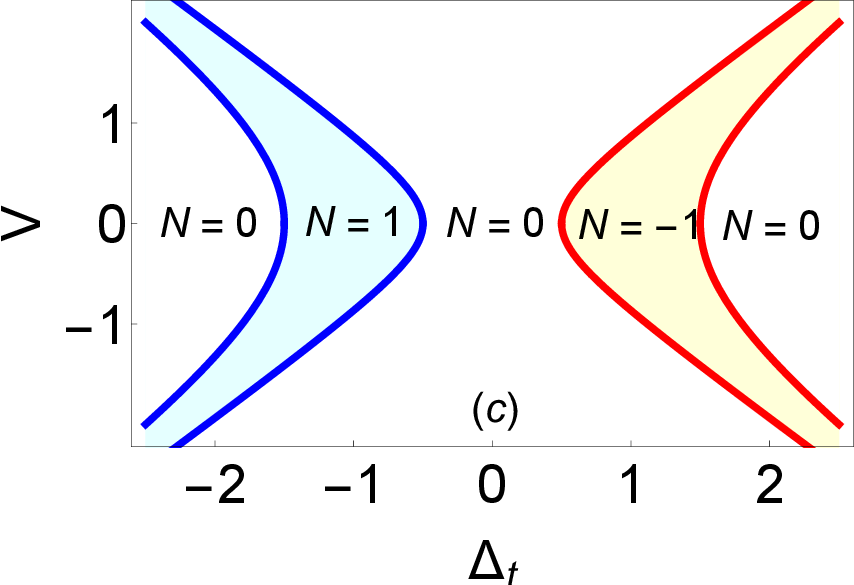}}
\end{minipage}
\begin{minipage}{0.23\textwidth}
	\centerline{\includegraphics[width=1\textwidth]{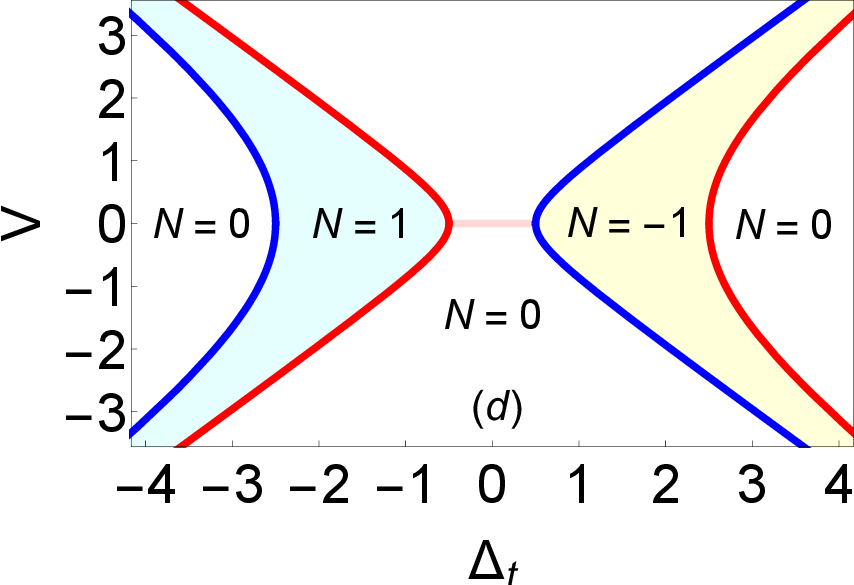}}
\end{minipage}
	\caption{
	{Topological phase diagrams in the presence of (a),(b) mirror symmetry breaking pairing interaction $\Delta_s$ and (c),(d) substrate potential $V$. The parameters are (a),(c) $\Delta=1$, $\mu=0$, $t=0.5$, and (b),(d) $t=1.5$. The $\chi$TSC phases bounded by the blue (red) curves are adiabatically connected to those of $H_D^+(H_D^-)$ at $V=0$ and $\Delta_s=0$. The pink lines in (b) at $V=0$ and (d) at $\Delta_s=0$ correspond to the mirror topological SC with  $N=0$ and $C=1$ shown in Fig.~\ref{fig3}(b).}
		\label{nu2}}
	\end{figure}

\subsection{Substrate potential}

{Substrates are usually present in thin-film experiments. The substrate potential effectively makes the chemical potential different on the top and bottom surfaces and breaks the inversion symmetry {${\cal I}=\sigma_0\chi_x$}. The structural inversion asymmetry (SIA) can thus be incorporated in the normal state Hamiltonian (\ref{h0}) by adding the SIA potential,
\begin{equation}
H_V=V\sigma_0\chi_z.
\label{sas}
\end{equation}
As a result, the substrate potential also breaks the mirror symmetry ${\cal M}$ in Eq.~(\ref{mirror}) of the TSS in the normal state, and ${\cal\widetilde M}$  in Eq.~(\ref{mirror-t}) of the BdG equation in the superconducting state. The modified BdG Hamiltonian in Eq.~(\ref{bdg}) with $V\neq0$ can no longer be block diagonalized, but continues to describe a nonunitary pairing state with broken TRS. To reveal the effects of the substrate potential, we obtain the topological phase diagrams shown in Figs.~{\ref{nu2}(c) and \ref{nu2}(d)} in the $\Delta_t-V$ plane. The phase structure is similar to the ones obtained for the mirror-symmetry breaking $\Delta_s$ shown in  Figs.~\ref{nu2}(a) and \ref{nu2}(b). In particular, the topological gap is robust and protects the chiral superconducting phase with nontrivial Chern numbers $N=1$ and $N=-1$, which are adiabatically connected to the $\chi$TSC at $V=0$ without a gap closing. Thus, despite mirror-symmetry breaking, the thin-film $\chi$TSC remains stable and belongs to class $D$ described by the nontrivial topological invariant $\mathbb{Z}$.}

\section{3D lattice model and quasi-2D ${\boldsymbol \chi}$TSC thin films \label{secVI}}

We now move beyond the effective theory of coupled superconducting TSS for the thin-film $\chi$TSC to discuss its materials realization in SCs with $Z_2$ nontrivial topological band structures. To this end, we construct a 3D cubic lattice model
capable of describing the superconducting TSS, which is motivated by Fe-based superconductor Fe(Te,Se). We demonstrate that as the number of layers along the $c$ axis is reduced to approach the quasi-2D thin-film limit, the nonunitary $\chi$TSC
with spontaneous TRS breaking can be induced by the coupling of the TSS in the top and bottom surfaces.
We note that a similar procedure has been used in the recently proposed time-reversal invariant TSC in thin-film Fe-based superconductors \cite{das}. We will discuss the differences in the physical approaches leading to the two kinds of intrinsic TSCs with or without TRS.
	
\begin{figure}
	\begin{minipage}{0.23\textwidth}
		\centerline{\includegraphics[width=1\textwidth]{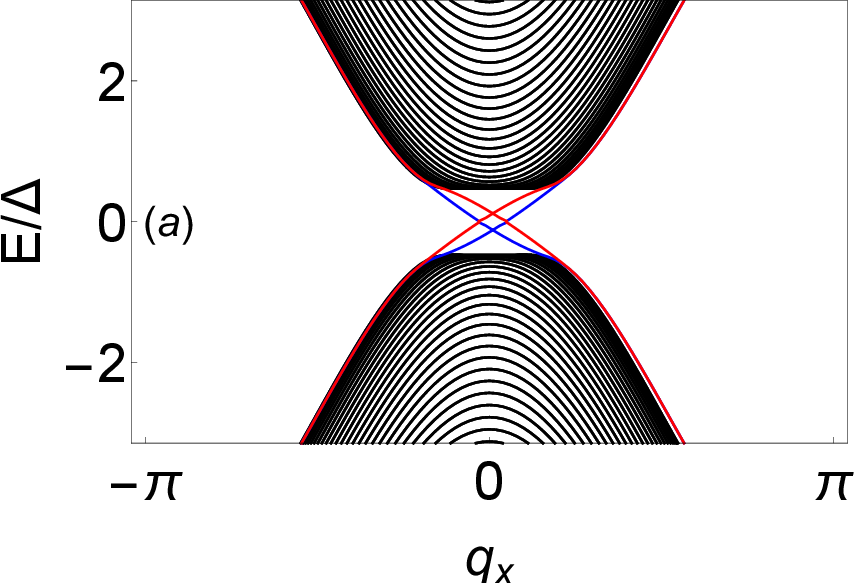}}
	\end{minipage}
	\begin{minipage}{0.23\textwidth}
		\centerline{\includegraphics[width=1\textwidth]{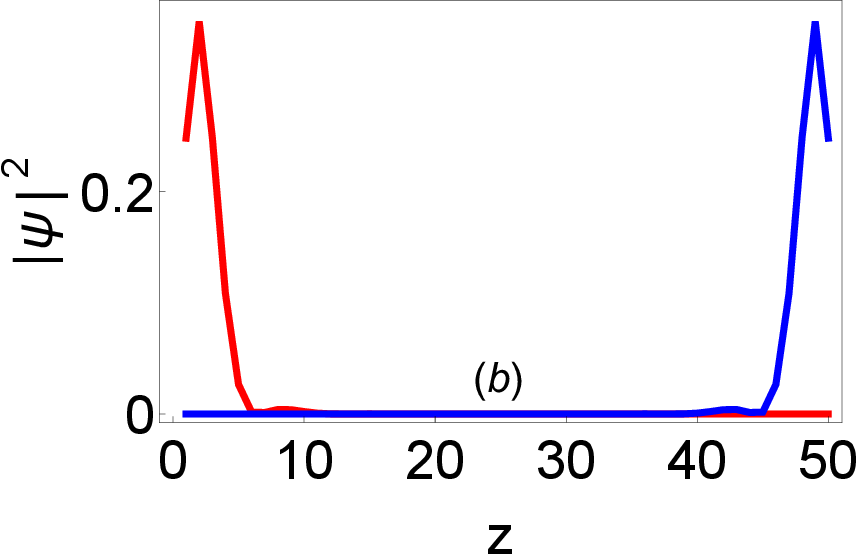}}
	\end{minipage}
	\begin{minipage}{0.23\textwidth}
		\centerline{\includegraphics[width=1\textwidth]{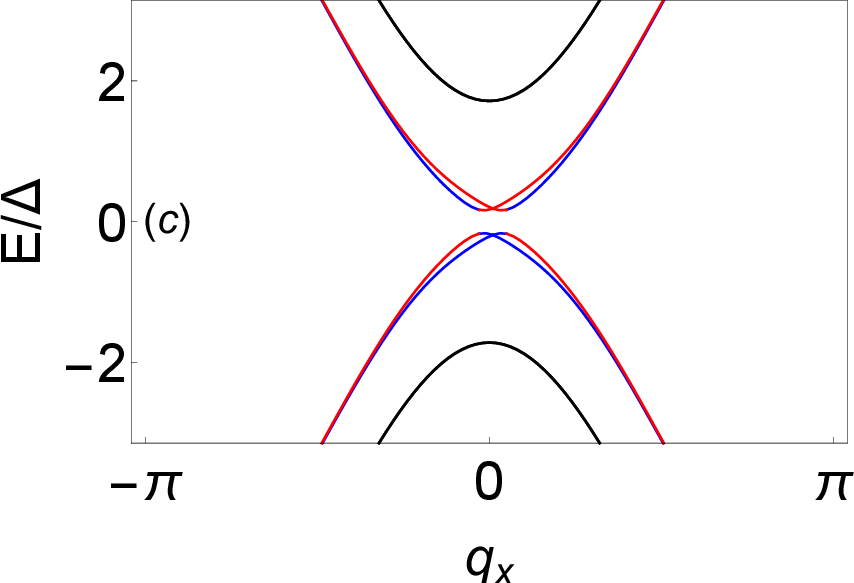}}
	\end{minipage}
	\begin{minipage}{0.23\textwidth}
		\centerline{\includegraphics[width=1\textwidth]{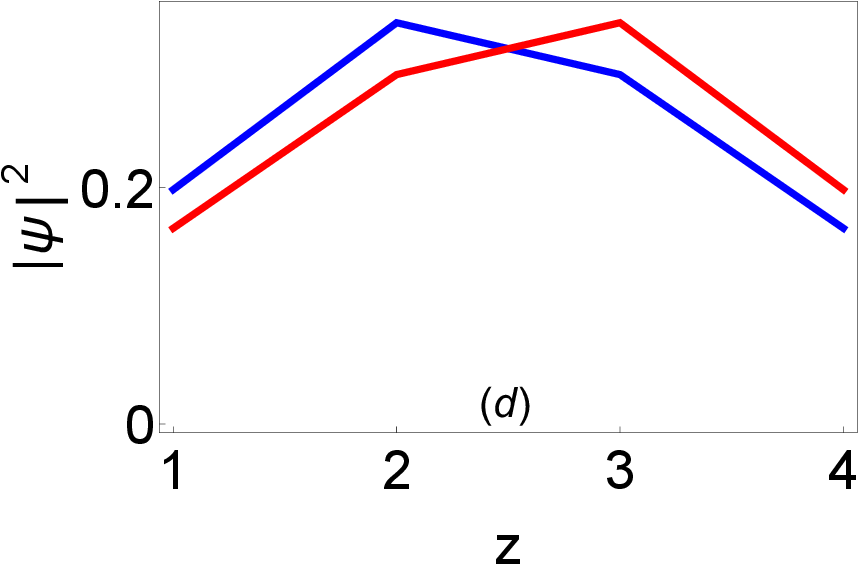}}
	\end{minipage}
	\caption{
{The band structure of the lattice model for the 3D TI with open boundaries along the $z$ direction plotted along the $q_x$ direction at $q_y=0$. The parameters are $(v,m_{0,1,2})=(1.0,-8.5,-3.0,3.0)$, and $V=0.5$ at the top and bottom surfaces. (a) The band dispersion in the bulk limit obtained with $n_z=50$ layers, showing the TSS split by the SIA potential $V$ into two Rashba Dirac cones (red and blue lines). (b) The spatial distribution of the in-gap states at the 2D $\Gamma$ point along the $z$ direction, which are independently localized on the top and bottom surfaces. (c) The thin-film limit band dispersion obtained with $n_z=4$ layers. An energy gap opens due to the coupling between the top and bottom weakly split Rashba Dirac cones with the corresponding spatial distribution of the states at the 2D $\Gamma$ point shown in (d).
}
		\label{nu}	}
\end{figure}

The minimal two-orbital and four-band 3D lattice model for a topological insulator is given by the Hamiltonian \cite{3dTI-cxliu},
\begin{eqnarray}
H_0&=&\sum_{\alpha=x,y,z} v \gamma_\alpha\sin q_\alpha +M({\bf q})\gamma_5 \label{tih} \\
M({\bf q})&=&[m_0-m_1(\cos q_x+\cos q_y)-m_2\cos q_z]\gamma_5,
\nonumber
\end{eqnarray}
where $\gamma_\alpha=\sigma_\alpha\rho_x$ and $\gamma_5=\sigma_0\rho_z$ with the Pauli matrices $\sigma_\alpha$ acting on the spin and $\rho_\alpha$ on the two bulk atomic orbitals, $v$ is the Fermi velocity, and $m_{0,1,2}$ are the band parameters. This model is known to be capable of producing a band inversion at the $Z=(0,0,\pi)$ point in the bulk Brillouin zone for $m_{0,1}< 0$, $m_2 >0$,  and $2m_1-m_2 < m_0 < \min(-m_2,-2m_1+m_2)$,
leading to a single helical Dirac cone around the $\Gamma$ point on the (001) surface \cite{das2}. This captures the basic features of the topological electronic structure of Fe(Te,Se) \cite{dxfz} and other Fe-based SCs observed by angle-resolved photoemission spectroscopy (ARPES) \cite{arpes-fts,pengzhang-natphys}.

To account for the substrate effect, we include the SIA potential $H_V=V\sigma_0\rho_z$
in the top and bottom layers of the lattice model and apply open boundary conditions along the $z$ direction.
Using  the parameters 
{$(v,m_{0,1,2})=(1.0,-8.5,-3.0,3.0)$} and a moderate {$V=0.5$},
we obtain the band structure for a thick film shown in Fig.~\ref{nu}(a) with $n_z=50$ number of layers.
The in-gap TSSs of the strong TI around the 2D $\Gamma$ point are well resolved and can be labeled by $|\nu\rangle$  as $|1\rangle\equiv|+\uparrow\rangle$,  $|2\rangle\equiv|+\downarrow\rangle$, $|3\rangle\equiv|-\uparrow\rangle$,  and $|4\rangle\equiv|-\downarrow\rangle$, where $\pm$ correspond to the weakly $V$-split Dirac cones.
%
%
%
The spatial distribution plotted in Fig.~\ref{nu}(b) shows that the TSSs are well localized at the top and bottom surfaces.
To simulate the thin-film limit, we reduce the layer thickness to $n_z=4$. The band dispersion of the thin film is plotted in Fig.~\ref{nu}(c) and the spatial distribution of the low-energy surface states in Fig.~\ref{nu}(d). There is significant mixing of the top and bottom surface states, leading to the massive Dirac fermion bands, which are weakly split by the substrate potential $V$.

Next, we consider the pairing interaction in the bulk. The pairing gap functions in Fe-based SCs have been observed to be ubiquitously three dimensional \cite{wang} and approximately follow $\Delta({\bf q})=\Delta_{2D}\cos q_x \cos q_y+\Delta_{\perp}(\cos q_x+\cos q_y)\cos q_z$. In the proposal for
the time-reversal invariant TSC in Fe-based SC thin films, $\Delta({\bf q})$ was taken to be two dimensional given by the first term, and the $s_\pm$ gap function changing sign in going from the zone center to the zone corner was crucial for producing a nodal ring in between the Fermi surfaces of the surface Rashba bands significantly split by a large SIA potential $V$  \cite{das}.
This should be contrasted to our proposal for the $\chi$TSC, where the physics takes place at small momenta around the bulk $Z$ and surface $\Gamma$ points. Thus motivated by Fe(Te,Se),  the bulk SC gap function in the lattice model can be approximated by the spin-singlet intraorbital pairing $\Delta^l i\sigma_y\rho_z$ in the long-wave length limit. This term is even under inversion $\mathcal{I}=\gamma_5$. Since inversion symmetry is broken by the SIA potential $V$, the inversion odd interorbital spin-singlet $\Delta_s^l i\sigma_y\rho_x$ and spin-triplet interorbital pairing $\Delta_t^l i \sigma_x \rho_y$ will be generated near the top and bottom layers. Thus, the pairing part of the Hamiltonian is given by
\begin{equation}
H_{\Delta}=\Delta^l  i\sigma_y\rho_z + \Delta_s^l i\sigma_y\rho_x +\Delta_t^l
i\sigma_x \rho_y,
\label{delta3D}
\end{equation}
where nonzero values of $\Delta_s^l$ and $\Delta_t^l $ are limited to the top and bottom two layers.
Note that $H_\Delta$ closely resembles the pairing Hamiltonian given in Eq.~(\ref{hsc2}) for the effective theory. The important difference is that the pairing is among the two atomic orbitals described by the Pauli matrices $\rho_\alpha$ in the 3D lattice model, whereas in the effective theory in Eq.~(\ref{hsc2}), pairing is among the electrons in the top and bottom surfaces described by the Pauli matrices $\chi_\alpha$.


Diagonalizing the full 3D lattice Hamiltonian $H_0+H_V+H_\Delta$, we obtain the spectra of the BdG quasiparticles shown in Fig.~\ref{fig9} for the bulk and the thin-film superconductors, using the same band parameters as in Fig.~\ref{nu} and $\Delta^l=1.3$, $\Delta_s^l= 1.5$, and $\Delta_t^l = 1.5$. Open boundary conditions are applied in both the $z$ and $y$ directions, such that in-gap states will appear when topological boundary modes are present.
Figure~\ref{fig9}(a) shows that the bulk superconductor ($n_z=20$) has a fully gapped spectrum and is not topological, absent of boundary states. Remarkably, an intrinsic $\chi$TSC emerges in the thin-film superconductor ($n=4$) where the coupling between the top and bottom surfaces enables TRS breaking as predicted by the effective theory. Figures~\ref{fig9}(b) and 7(c) reveal that the thin-film superconductor has a spectrum with continuous in-gap states due to topological boundary excitations. The latter correspond to the $\chi$MEM localized on the physical edge of the thin-film superconductor in the $y$ direction, as shown in the spatial distribution in Fig.~\ref{fig9}(d), consistent with the effective theory.
	
	\begin{figure}
		\begin{minipage}{0.23\textwidth}
			\centerline{\includegraphics[width=1\textwidth]{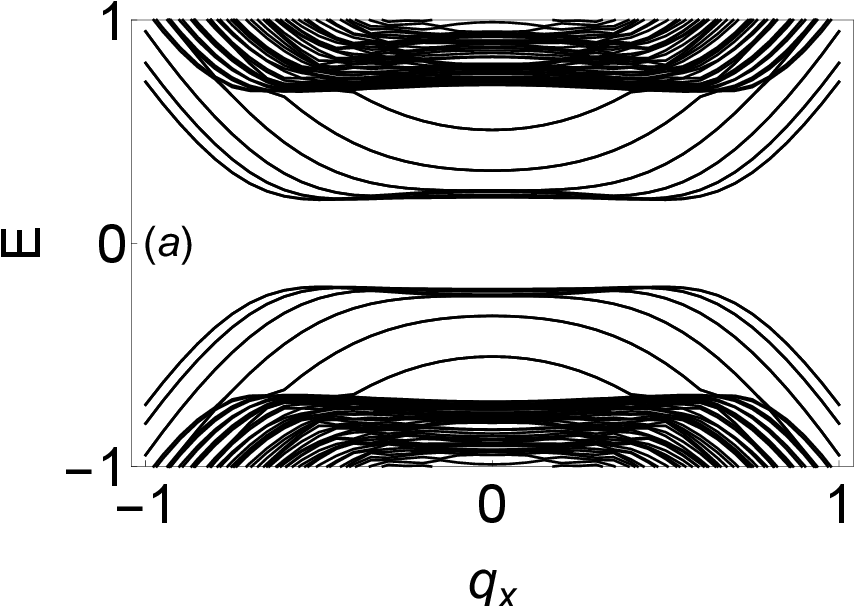}}
		\end{minipage}
		\begin{minipage}{0.23\textwidth}
			\centerline{\includegraphics[width=1\textwidth]{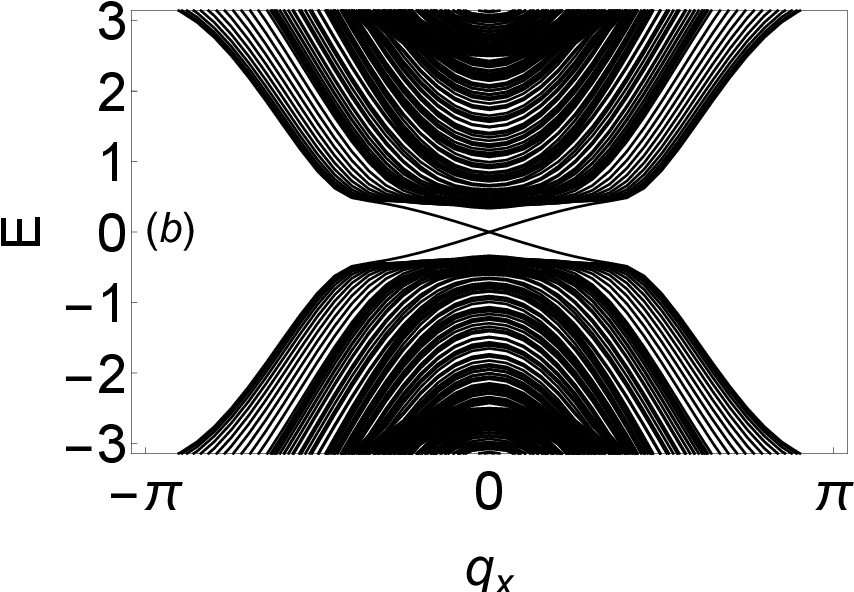}}
		\end{minipage}
		\begin{minipage}{0.23\textwidth}
			\centerline{\includegraphics[width=1\textwidth]{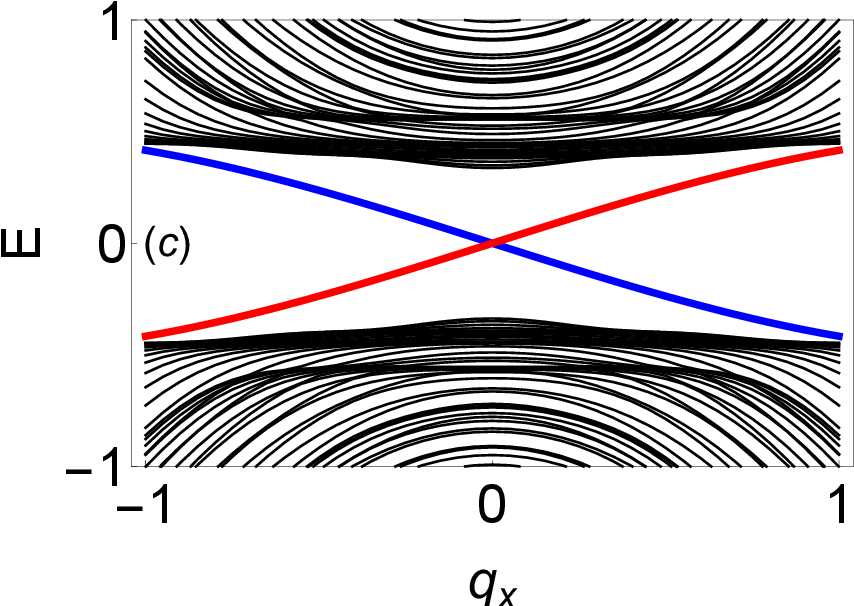}}
		\end{minipage}
		\begin{minipage}{0.23\textwidth}
			\centerline{\includegraphics[width=1\textwidth]{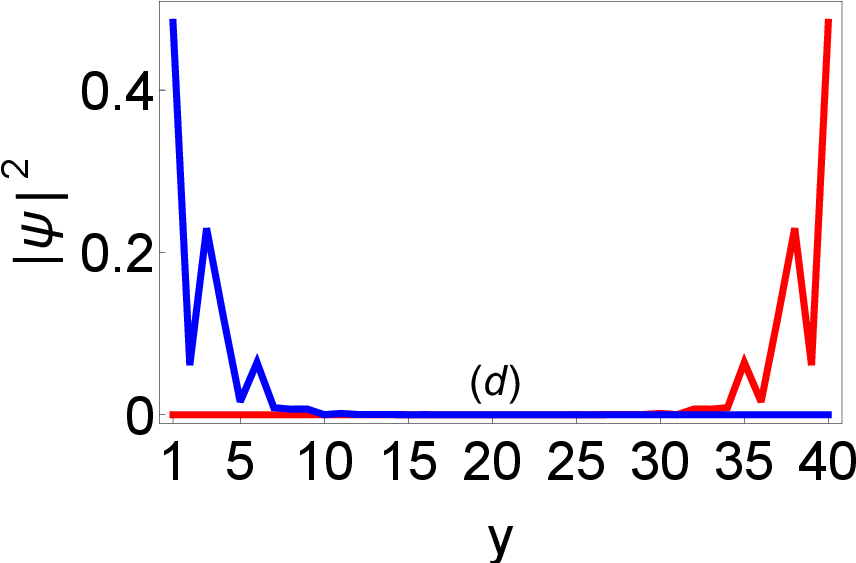}}
		\end{minipage}
		\caption{
Energy spectra of BdG quasiparticles in the SC state of the 3D lattice model with the same band parameters as in Fig.~(\ref{nu}). Open boundary conditions are applied in the $z$ and $y$ directions for $n_z$ number of layers of width $L_y$. The $x$ direction is periodic with continuous momentum $q_x$. The pairing parameters are $\Delta^l = 1.3$, $\Delta_s^l = 1.5$, and $\Delta_t^l = 1.5$.
(a) The spectrum for $n_z=20$ and $L_y=20$ is fully gapped, consistent with a topologically trivial bulk superconductor. (b) The spectrum for $n_z=4$ layers and $L_y=40$ shows a SC gap with in-gap states, realizing a thin film $\chi$TSC with $\chi$MEMs.
(c) Zoom-in of (b). The blue and red curves represent the $\chi$MEMs.
(d) The spatial distribution of the layer-averaged wave functions of the $\chi$MEMs at an excitation energy $E=0.05$ in (c), showing localization at the two edges in the $y$ direction.
			\label{fig9}	}
	\end{figure}


\section{Detecting half-integer quantized conductance on an antidot device \label{secVII}}

 The half-quantized conductance $\frac{e^2}{2h}$ is a unique transport signature of a single {$\chi$MEM} and provides a crucial test for the $\chi$TSC \cite{science,hih}. We propose here a device made of the thin film SC patterned with an antidot, as shown in Fig.~\ref{fig4s}. The chemical potential inside the antidot can be adjusted by adding a gate to this area [{Fig. \ref{fig4s}(a)}]. We next show that there is a $\chi$MEM propagating along the edges of the antidot, which can be observed by directly measuring the two-terminal conductance in a weak external magnetic field [Fig.~\ref{fig4s}b]. Without loss of generality, we consider $N_+=1$ and $N_-=0$, where the low-energy physics of $H_D^+$ is sufficiently described by the two bands closest to the Fermi energy in the effective BdG equation (see {Appendix \ref{app3}}),
\begin{equation}
\left(
\begin{array}{cccc}
-M_1  & D_x-iD_y \\
D_x+iD_y & M_1
\end{array}
\right)\left(\begin{array}{c}u\\ v\end{array}\right)=0, \label{ebdg}
\end{equation}
where $D_\alpha=-i(\partial_a+iA_\alpha)$ is the covariant derivative, $A_\alpha$ the vector potential, {$M_1=t- \sqrt{(\Delta+\Delta _t) ^2+\mu ^2}$,} and $u$ and $v$ the wave functions of the Bogoliubov quasiparticle. $\nabla\times {\bf A}={\bf B}$  gives the weak magnetic field applied perpendicular to the thin-film SC. The Andreev reflection is assumed to be suppressed, so the only conducting channel is along the edge.  Near the edge of the antidot, {$M_1\equiv\tilde{\mu}(x,y)$} is negative on the antidot side, with amplitudes much smaller than the positive bulk values of $M_1$ on the $\chi$TSC side. This ensures that the strength of the magnetic field $B$ does not exceed the lower critical field of the thin-film SC. For simplicity, we take $\tilde\mu$ to be a constant and assume the magnetic field $B$ to be uniform on the order of the penetration length. In the numerical calculations, we use $B=0.5\vert\tilde\mu\vert$ and $\vert\tilde\mu\vert\sim10^{-3}\Delta$.

Diagonalizing Eq.~(\ref{ebdg}) in real space, the spectrum of the gapless chiral edge states can be obtained, which includes the zero-energy bound state carrying a vanishing momentum in the direction parallel to the edges. In Fig.~\ref{fig4s} (c), the spatial distributions of $|u|$ and $|v|$ are plotted for the zero-energy bound state. For $B=0$, $\vert u\vert=\vert v\vert$, and they are localized at both edges (red curves), as expected for a charge-neutral chiral Majorana edge state which cannot be detected directly by charge transport. Applying the weak magnetic field $B\ne0$, we find that while $\vert u\vert$ (blue curve) is localized on one edge, $\vert v\vert$ (blue curve) is localized on the opposite edge, as shown in Fig.~\ref{fig4s}(c). Therefore, the chiral Majorana edge states have been transformed into Jackiw-Rebbi-type solitons \cite{JR}. As a result, the chiral charge current can now be carried by a quasielectron current with probability $|u|^2$ on one edge and a quasihole current with probability $|v|^2$ on the opposite edge. For the device shown in {Fig. \ref{fig4s}(b)}, the Landauer-B\"{u}ttiker formula for the ballistic currents $I_1$ and $I_2$ are given by \cite{LB}
\begin{equation}
I_1=-I_2=\frac{e^2}{h}(|u|^2V_1-|v|^2V_2).
\end{equation}
Choosing $V_1=-V_2=V/2$, the current-voltage relation becomes $I_1=-I_2=\frac{e^2}{h}(|u|^2+|v|^2)\frac{V}{2}=\frac{e^2}{2h}V$, giving rise to the half-quantized conductance $\frac{e^2}{2h}$.

\begin{figure}
	\includegraphics[width=0.45\textwidth]{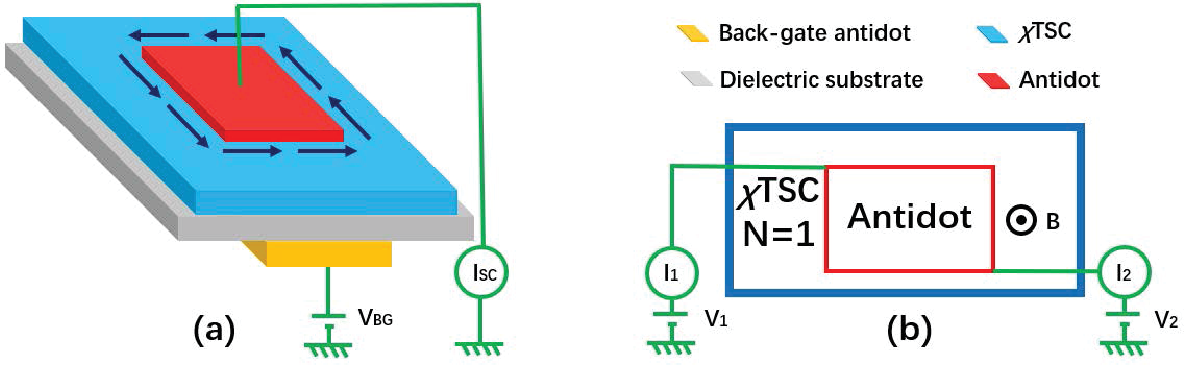}
	\centerline{}
	\begin{minipage}{0.45\textwidth}
		\centerline{\includegraphics[width=0.8\textwidth]{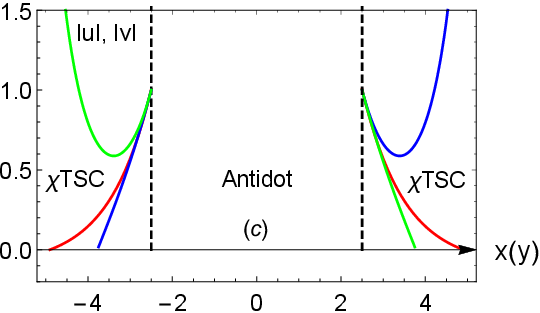}}
	\end{minipage}
	\caption{
		{(a) Schematics of the device for detecting the half-quantized conductance of a single $\chi$MEM at the boundaries of the $\chi$TSC. The red, blue, brown, and orange layers are the antidot, $\chi$TSC, dielectric substrate and back-gate antidot, respectively. The arrows indicate the chiral $\chi$MEM. (b) The top view of the device in (a). An external magnetic field $B$ is applied. $V_1$ and $V_2$ are the voltages applied to leads 1 and 2. (c) The spatial distributions of $|u|$ and $|v|$ for the zero-energy bound state. The red curves are for $|u|$ and $|v|$ when $B=0$; both are localized at the edges. For {$B/\tilde{\mu}=0.5$} and in the Landau gauge, $|u|$ (blue) is localized while $|v|$ (green) merges into the bulk at one edge, and $|v|$ is localized and $|u|$ merges into the bulk at the opposite edge.  \label{fig4s}	}}
\end{figure}
\section{Discussions\label{secVIII}}

The mirror symmetry of the coupled TSSs
is a key ingredient in our theory for the spontaneous TRS breaking and chiral topological superconductivity in thin-film SCs with a ${\mathbb Z}_2$ nontrivial topological band structure. Under the mirror symmetry, the superconducting state is described by two noncollinear and spatially isotropic ${\mathbf d}$ vectors for spin-triplet and orbital-triplet pairing, i.e.,
a nonunitary 2D $\chi$TSC with spontaneous TRS breaking and a nonzero spin-orbit polarization. As a result, the thin-film $\chi$TSC hosts a single $\chi$MEM at the boundary and a MZM in the vortex core. This is in contrast to the time-reversal invariant TSC with momentum-dependent pairing and rotational symmetry breaking proposed for a Cu intercalated bulk topological insulator Cu$_x$Bi$_2$Se$_3$ \cite{fuberg,fu} and thin films \cite{Luka}, and the chiral  $p$-wave TSC proposed for the interface of noncentrosymmetric Pb$_3$Bi/Ge(111) \cite{wlz}.
The intrinsic $\chi$TSC remains stable in the presence of mirror-symmetry breaking interactions, such as 
{the substrates and} the spin-singlet pairing between the top and bottom surfaces \cite{sato-prb}, and belongs to class $D$ characterized by a topological invariant $\mathbb{Z}$.

The $\chi$TSC thin films can be potentially materialized using the Fe-based SCs.
The Fe-chalcogenides and pnictides are $p$-$d$ charge transfer metals in the normal state \cite{zhoukotliarwang}. A $p$-$d$ band inversion can develop and
produce a $\mathbb{Z}_2$ topological metal with lightly doped TSSs \cite{dxfz,jphu,gangxu}. The latter have been observed by spin-polarized angle-resolved photoemission spectroscopy in bulk Fe(Te,Se) and LiFeAs \cite{arpes-fts,pengzhang-natphys}.
Below the bulk $T_c$, the TSSs become superconducting with an energy gap comparable to the bulk gap \cite{arpes-fts,gangxu}.
Candidate MZMs have been observed inside magnetic-field-induced vortices in FeTe$_{0.55}$Se$_{0.45}$ \cite{dinggao2018,hanaguri}, (Li$_{0.84}$Fe$_{0.16}$)OHFeSe \cite{donglai-prx}, CaKFe$_4$As$_4$ \cite{CaKFe4As4}, and the quantum anomalous vortices \cite{qav} nucleated at excess and adatom magnetic Fe sites in FeTe$_{0.55}$Se$_{0.45}$ \cite{yin-natphys2015,qav-mzm2020} and LiFeAs \cite{yin2019} without applying external magnetic fields. Our effective theory and the 3D lattice model calculations
clearly show that, in the absence of coupling between the top and bottom surfaces, the superconducting TSSs do not constitute a topological superconductor and there are no gapless boundary excitations.
The chiral topological superconductivity proposed here can be potentially realized by fabricating thin-film SCs using this class of Fe-based SCs to allow the coupling of TSSs on the top and bottom surfaces.
Growing Fe(Te,Se) thin-film SCs with controlled thickness by the molecular beam epitaxy (MBE) method or fabricating Fe(Te,Se) films by exfoliation are realistic under the current experiential techniques. Other materials choices include the noncentrosymmetric PbTaSe$_2$ where superconducting TSSs have been observed \cite{guan2016,chang2016,yin2019} and the recently discovered  kagome lattice SCs,
$A$V$_3$Sb$_5$ ($A$=K, Rb, Cs), with a $Z_2$ topological band structure \cite{wilson-prm,wilson-prl,proximity}.

In addition to realizing a different and profound topological quantum state of matter,
the chiral topological SC thin-films provide an unparalleled single-material platform for not only non-Abelian vortex MZMs, but also robust $\chi$MEMs for designing topological logic gates.
The detection of the half-quantized conductance and the non-Abelian braiding of the $\chi$MEMs using the $\chi$TSC thin-film-based devices can be more feasible than the previously proposed QAHI-SC proximity effect hybrid structures.
Supported by the scalability of the $\chi$MEMs in arrays of patterned antidots as well as superlattice stacking, the $\chi$TSC thin-films hold promise for the exploration of fault-tolerant quantum computing.

\acknowledgements

We thank Bin Chen, Dong-Lai Feng,  Zheng-Cheng Gu, Kun Jiang, Jing Wang, Yi-Hua Wang, Yong-Shi Wu, Yi Zhang, and Sen Zhou for helpful discussions. This work is supported in part by NNSF of China with Grants No. 12174067 (XL,YY), No. 11474061 (XL,YGC,YY) and No. 11804223 (XL). Z.W. is supported by the U.S. Department of Energy, Basic Energy Sciences Grant No. DE-FG02-99ER45747.

\appendix

\section{Time-reversal symmetry \label{app1}}

\begin{table}[htb]
\caption{Summary of the time-reversal operators for Hamiltonian (\ref{bdg}) in the untransformed Nambu basis. $\pm$ represent whether or not the corresponding term is time-reversal invariant.}
	\begin{center}
		\begin{tabular}{|c|c|c|c|c|c|c|c|}
			\hline $\mathcal{T}$ & $\mathcal{T}^2$ & $\mu$-term & $\Delta$-term& $t$-term & $\Delta_t$-term & $\lambda$-term \\
			\hline $i\sigma_y\chi_0\tau_0\mathcal{K}$ & $-1$ & $+$ & $+$ & $+$ & $-$ & $-$ \\
			\hline $i\sigma_y\chi_0 \tau_z\mathcal{K}$ & $-1$  & $+$ & $-$ & $+$ & $+$ & $-$ \\
			\hline $ i\sigma_y\chi_z\tau_0\mathcal{K}$ & $-1$  & $+$ & $+$ & $-$ & $+$ & $-$ \\
			\hline
		\end{tabular}
	\end{center} \label{table1}
\end{table}

Time-reversal invariance is restored when one of the three parameters $t$, $\Delta$, and $\Delta_t$ vanishes. Specifically, consider the case where
$\Delta_t=0$. Then $(i\sigma_y\tau_0\mathcal{K})H_D^+ ({\bf q})(i\sigma_y\tau_0\mathcal{K})^{-1}=H_D^-(-{\bf q})$. The time-reversal operator for $H_D^+\oplus H_D^-$ is therefore given by $\mathcal{T}_{\Delta_t=0}=i\sigma_y\tau_0\Theta_x\mathcal{K}$, where $\Theta_i$ are Pauli matrices acting on the mirror index. In the original Nambu basis $\Psi_N$, one finds $U^{-1}\mathcal{T}_{\Delta_t=0}U=i\sigma_y\chi_0\tau_0\mathcal{K}$, which is the usual time reversal operator $\mathcal{T}$. Thus, when $\Delta_t=0$, the Hamiltonian (\ref{bdg}) describes a time-reversal invariant TSC \cite{TI-review}.
Next, consider $\Delta=0$, so that $(\sigma_x\tau_z\mathcal{K})H_D^+({\bf q})(\sigma_x\tau_z\mathcal{K})^{-1}=H_D^-(-{\bf q})$. This allows us to define the modified time-reversal operator $\mathcal{T}_{\Delta=0}=i\sigma_x\tau_z\Theta_y\mathcal{K}$. In the original $\Psi_N$ basis, it becomes $U^{-1}\mathcal{T}_{\Delta=0}U=-i\sigma_y\chi_0\tau_z\mathcal{K}=-\tau_z\mathcal{T}$, i.e., the usual time-reversal followed by a discrete operation in the particle-hole sector. Finally, if $t=0$, we have $(i\sigma_y\tau_x\mathcal{K})H_D^\pm
({\bf q})(i\sigma_y\tau_x\mathcal{K})^{-1}=H_D^\pm (-{\bf q})$. As a result, each of $H_D^+$ and $H_D^-$ has TRS. The operators $\mathcal{T}, \mathcal{T}_{\Delta_t=0}$, and $\mathcal{T}_{\Delta=0}$ are all possible time-reversal operators that square to $-1$. {We summarize the results in Table \ref{table1}.} When none of the three parameters $t$, $\Delta$, and $\Delta_t$ is zero, the TRS is spontaneously broken in the BdG Hamiltonian (\ref{bdg}).

\medskip

\section{lattice model \label{app2}}

 \begin{figure}
	\begin{minipage}{0.23\textwidth}
		\centerline{\includegraphics[width=1\textwidth]{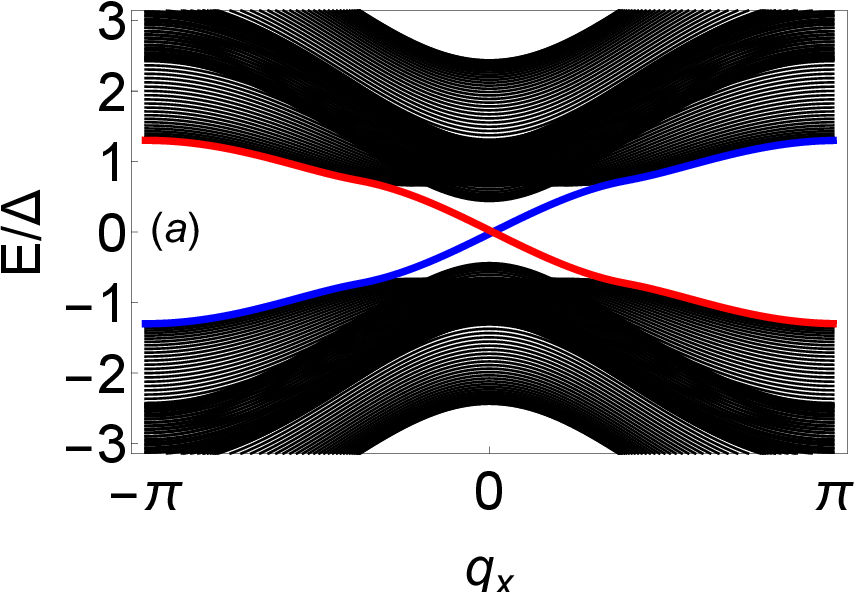}}
	\end{minipage}
	\begin{minipage}{0.23\textwidth}
		\centerline{\includegraphics[width=1\textwidth]{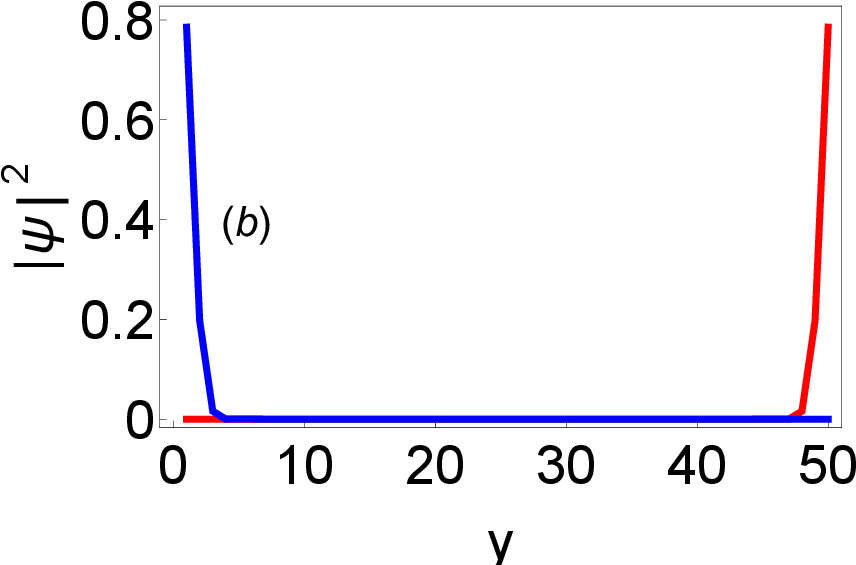}}
	\end{minipage}
	\begin{minipage}{0.23\textwidth}
		\centerline{\includegraphics[width=1\textwidth]{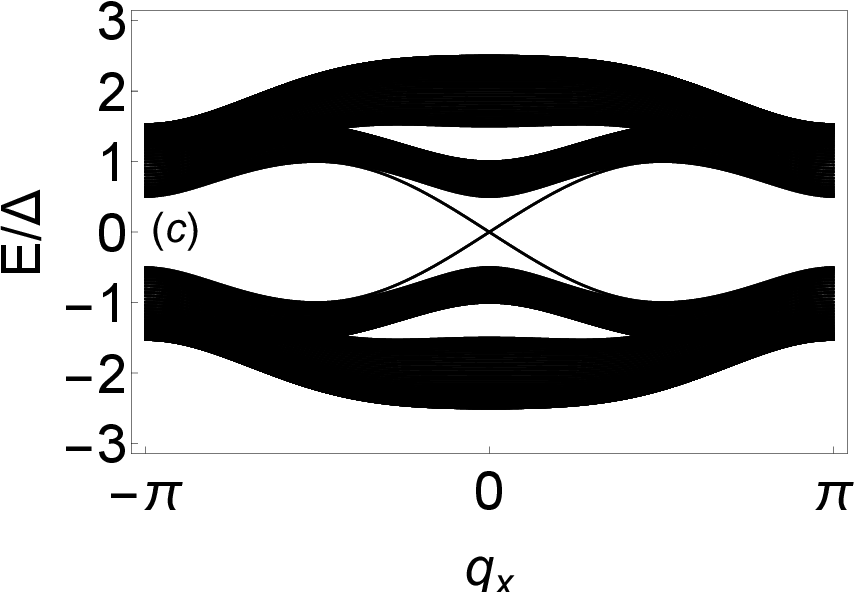}}
	\end{minipage}
	\begin{minipage}{0.23\textwidth}
		\centerline{\includegraphics[width=1\textwidth]{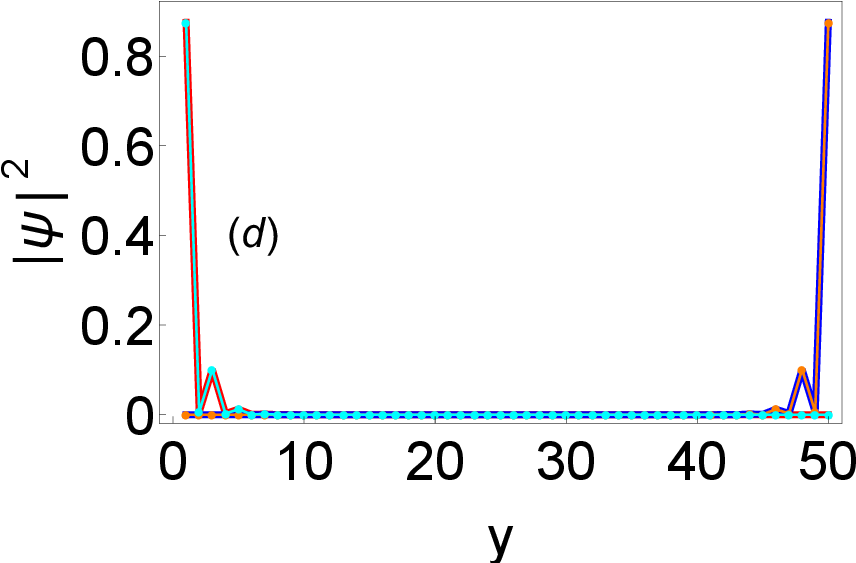}}
	\end{minipage}
	\caption{
		(a),(c) Energy spectrum of the lattice model under open boundary conditions in the $y$ direction for $\Delta=1$. (a) $\mu=2$, $t_0=2.5$, $t_1=1.0$, $\Delta_t=1.5$; (c) $\mu=0$, $t_0=1.5$,  $t_1=-0.5$; and $\Delta_t=0$.
(b),(d) Spatial distributions of the gapless edge states in (a) and (c) along the $y$ direction (50 sites) at a fixed energy $E_a=0.37$ and $E_c=0.48$, respectively. The red and blue curves in (b) are the gapless chiral edge states of a single $\chi$MEM. In (d), there are two counter propagating $\chi$MEMs, namely, a helical Majorana edge mode at each boundary. They are represented by the solid and dotted lines.
		\label{nu1}	}
\end{figure}

 To test the topological SC phases of the BdG Hamiltonian (\ref{bdg}), we study a tight-binding model on the square lattice with periodic boundary condition along the $x$ direction and open boundary condition along the $y$ direction and obtain the chiral Majorana edge states directly along the open boundary. Specifically, we replace the hopping amplitude $t\rightarrow t_0+t_1(q_x^2+q_y^2)$ in Eq.~ (\ref{bdg}) and discretize the momenta $q_{x,y}\rightarrow a^{-1}\sin{(q_{x,y}a)}$ and $q_x^2+q_y^2\rightarrow 2-2a^{-2}(\cos(q_xa)+\cos(q_ya))$, where $a$ is the lattice constant (set to unity). We numerically diagonalize the lattice BdG Hamiltonian to obtain the eigenstate spectrum and the spatial distributions of the gapless edge modes along the open boundaries on lattices with 50 sites in the $y$ direction. The results are shown in Fig.~\ref{nu1}, where the eigenstate spectra in Figs.~\ref{nu1}(a) and \ref{nu1}(c) have been shown in Fig.~\ref{fig1}(f) and Fig.~\ref{fig3}(d) in the main text and reproduced for comparison. The corresponding distributions of the Majorana edge modes are shown in Figs.~\ref{nu1}(b) and \ref{nu1}(d), respectively.

\medskip

\section{Ginzburg-Landau theory\label{app-gl}}

{We consider the Ginzburg-Landau theory to analyze the relative phase between the intraorbital-triplet spin-singlet pairing $\Delta$ and the interorbital-singlet spin-triplet pairing $\Delta_t$. First, we consider the symmetry properties of the pairing terms of the superconducting TSS of the thin-film SC. As mentioned in the main text, there is a mirror symmetry $\mathcal{M}=-i\sigma_z\chi_x$ due to the opposite helicities of the Dirac fermion TSSs and, in general, the superconducting pairing order parameter ${\bf \Delta}$ takes the following form:
	\begin{equation}
		{\bf \Delta}=\Delta_\alpha (i\sigma_y)({\bf d^\prime}\cdot{\boldsymbol\chi}) (i\chi_y)+\Delta_{t\beta} ({\bf d}\cdot{\boldsymbol\sigma})(i\sigma_y) (i\chi_y),
		\label{delta-gl}
	\end{equation}
where $\sigma(\chi)$ labels the spin(surface/orbital) index, $({\bf d^\prime})$ and $({\bf d})$ are two unit triplet $d$ vectors in the orbital and spin
 sectors, and $\alpha$ and $\beta$ correspond to the $(x,y,z)$ indices of the $({\bf d^\prime})$ and $({\bf d})$ components. Here we focus on the momentum independent pairings because they are favored over the momentum-dependent pairings in the weak coupling limit with purely short-range interactions \cite{nagaosa-prl}. In this case, the six pairing terms can be grouped in two parts under mirror symmetry $\mathcal{M}$, namely, $\mathcal{M}	{\bf \Delta}\mathcal{M}^{-1}=	{\bf \Delta}$ for $\Delta_x$ and $\Delta_{tz}$ (or  ${\bf d^\prime}=\hat x$ and ${\bf d}=\hat z$), while  $\mathcal{M}	{\bf \Delta}\mathcal{M}^{-1}=	-{\bf \Delta}$ for the other four terms. Furthermore, we have chosen a square lattice geometry for the lattice model with $D_{4h}$ symmetry; then, $\Delta_x$, $(\Delta_y,\Delta_z)$, $(\Delta_{tx},\Delta_{ty})$, and $\Delta_{tz}$ belong to the irreducible representations $A_{2u}$, $A_{1g}$, $E_{u}$, and $A_{1u}$ respectively \cite{nagaosa-prl}. We summarize the above properties in Table \ref{table2}.}

\begin{table}[htb]
	\caption{Summary of the symmetry properties of the pairing term ($\ref{delta-gl}$). $\Psi_i$ is the fermion annihilation operator with $i$ being the surface/orbital index. The parity operation is defined by $(x,y,i)\rightarrow(-x,-y,\bar{i})$ with the exchange of surface/orbital index $i\rightarrow \bar{i}$ \cite{nagaosa-prl}. Here $+(-)$ means even (odd). }
	\begin{center}
		\begin{tabular}{|c|c|c|c|c|c|c|c|}
			\hline pairing terms & mirror $\mathcal{M}$ & irrep. of $D_{4h}$ & parity in $D_{4h}$ \\
			\hline $\Delta_x:(\psi_{1\uparrow}\psi_{1\downarrow}-\psi_{2\uparrow}\psi_{2\downarrow})$ & $+$ & $A_{2u}$ & $-$  \\
			\hline $\Delta_y:i(\psi_{1\uparrow}\psi_{1\downarrow}+\psi_{2\uparrow}\psi_{2\downarrow})$ & $-$  & $A_{1g}$ & $+$ \\
			\hline $ \Delta_z:(\psi_{1\uparrow}\psi_{2\downarrow}-\psi_{1\downarrow}\psi_{2\uparrow})$ & $-$  & $A_{1g}$ & $+$ \\
			\hline $\Delta_{tx}:(\psi_{1\uparrow}\psi_{2\uparrow}-\psi_{1\downarrow}\psi_{2\downarrow})$ & $-$ & $E_u$ & $-$  \\
			\hline $\Delta_{ty}:i(\psi_{1\uparrow}\psi_{2\uparrow}+\psi_{1\downarrow}\psi_{2\downarrow})$ & $-$  & $E_u$ & $-$  \\
			\hline $ \Delta_{tz}:(\psi_{1\uparrow}\psi_{2\downarrow}+\psi_{1\downarrow}\psi_{2\uparrow})$ & $+$  & $A_{1u}$ & $-$  \\
			\hline
		\end{tabular}
	\end{center} \label{table2}
\end{table}

{In the following, we will focus on the Ginzburg-Landau theory of the mirror-even pairing terms $\Delta_x$ and $\Delta_{tz}$. For the thin-film geometry which lacks the full three-dimensional inversion symmetry, the mixing between spin-singlet and spin-triplet pairings is allowed \cite{RPP2017,JPCM2021}. Then the fourth-order invariant of the Ginzburg-Landau free energy can be approximated as \cite{JPCM2021,AP1990,RMP1991},
\begin{equation}
	f_4=\gamma_1(|\Delta_{x}|^2+|\Delta_{tz}|^2)^2+\gamma_2|\Delta_{x}^2+\Delta_{tz}^2|^2+\gamma_3(|\Delta_{x}|^4+|\Delta_{tz}|^4),
\end{equation}
where $\gamma_1,\gamma_2,\gamma_3$ are materials dependent parameters. For simplicity, we choose, $\Delta_{x}=|\Delta|\cos\theta$ and $\Delta_{tz}=e^{i\omega }|\Delta|\sin\theta $, and the above free energy can be rewritten as
\begin{equation}
	f_4=|\Delta|^4(\gamma_1+\gamma_2+\gamma_3-\sin^2(2\theta)(\gamma_3-2\gamma_2\sin^2(2\omega))/2).
\end{equation}
Minimizing $f_4$ with respect to $\theta$ and $\omega$ reads
\begin{equation}
	\sin(4\theta)(\gamma_3-2\gamma_2\sin^2\omega)=0,\quad \gamma_2\sin^2(2\theta)\sin(2\omega)=0,
\end{equation}
with three nontrivial sets of solutions, $(\Delta_x,\Delta_{tz})\propto(\pm 1,0)$, $(\pm 1,\pm 1)$, and $(\pm 1,\pm i)$. $(\pm 1,\pm 1)$ breaks time-reversal symmetry as discussed in the main text, while the other two cases preserves time-reversal symmetry.}

\medskip

\section{Two-band Hamiltonian \label{app3}}
Here, we derived the effective two-band BdG Hamiltonian in Eq.~(\ref{ebdg}). Without loss of generality, we consider the topological nontrivial phase of the $4\times 4$ BdG Hamiltonian $H_D^+$ in Eq.~(\ref{hpm}),
\begin{equation}
	H_D^+=\left(
	\begin{array}{cccc}
		\mu -t & \Delta +\Delta _t & 0 & q_x+i q_y \\
		\Delta +\Delta _t & -t-\mu  & q_x+i q_y & 0 \\
		0 & q_x-i q_y & t-\mu  & -\Delta -\Delta _t \\
		q_x-i q_y & 0 & -\Delta -\Delta _t & t+\mu  \\
	\end{array}
	\right).
\nonumber
\end{equation}
The spectrum of the Hamiltonian can be obtained analytically and is given by the following four energy bands,
\begin{eqnarray}
	E_{\pm,\pm}({\bf q})&=&\pm\sqrt{{\bf q}^2+t^2+\omega^2 \pm2 \sqrt{\mu ^2 {\bf q}^2+t^2 \omega^2}  }\nonumber\\
	=&\pm&\sqrt{(1-\frac{\mu ^2}{ \omega t}){\bf q}^2
		+(t\pm \omega)^2+\mathcal{O}({\bf q}^4)
	 },
\end{eqnarray}
where ${\bf q}^2=q_x^2+q_y^2$ and $\omega^2=(\Delta+\Delta_t)^2+\mu^2$,
and we have expanded ${\bf q}^2$ to the lowest order near the $\Gamma$ point in the second line. In the long-wave-length limit, the spectrum is indeed relativistic. The mass term for the two bands closest to the Fermi energy is given by $M_1=t- \sqrt{(\Delta+\Delta _t) ^2+\mu ^2}$. When $\mu\ll |t(\Delta+\Delta_t)|$, the effective two-band BdG Hamiltonian reads,
\begin{equation}
H_{eff}=	\left(
	\begin{array}{cccc}
		-M_1  & q_x-iq_y \\
		q_x+iq_y & M_1
	\end{array}
	\right),
\end{equation}
which is equivalent to a chiral $p+ip$ superconductor.

\end{document}